\documentclass[a4paper,fleqn,usenatbib]{mnras}

\usepackage{newtxtext,newtxmath}

\usepackage[T1]{fontenc}
\usepackage{ae,aecompl}
\usepackage{float}

\usepackage{graphicx}	
\usepackage{amsmath}	
\usepackage{amssymb}	
\usepackage{mathrsfs} 
\usepackage{natbib}
\usepackage{amsmath}
\usepackage{float}
\usepackage{amssymb}
\usepackage{color}
\usepackage{captcont}
\usepackage{url}
\usepackage{rotating}
\usepackage{longtable,array}
\usepackage{afterpage}
\usepackage{pdflscape}
\usepackage{threeparttablex}
\usepackage{booktabs} 
\usepackage{captcont}
\usepackage{tabularx}

\newcommand\degrees[1]{\ensuremath{#1^\circ}}
\newcommand{\as}{$^{\prime\prime}$}
\newcommand{\MA}[1]{}

\title[Dipper Disc Inclinations]{
\LARGE Are Inner Disc Misalignments Common? ALMA Reveals an Isotropic Outer Disc Inclination Distribution for Young Dipper Stars}

\author[M. Ansdell et al.]{ \Large M. Ansdell,$^{1,2,3}$\thanks{E-mail: mansdell@flatironinstitute.org} E. Gaidos,$^{4}$ C. Hedges,$^{5}$ M. Tazzari,$^{6}$ A. L. Kraus,$^{7}$ M. C. Wyatt,$^{6}$ G. M. Kennedy,$^{8, 9}$ \newauthor \Large J. P. Williams,$^{10}$ A. W. Mann,$^{11}$ I. Angelo,$^{2, 12}$ G. D\^{u}chene,$^{2, 13}$ E. E. Mamajek,$^{14, 15}$ J. Carpenter,$^{16}$ \newauthor \Large T. L. Esplin,$^{17}$ A. C. Rizzuto$^{7}$ \\ \\
$^{1}$Center for Integrative Planetary Science, University of California at Berkeley, Berkeley, CA 94720, USA\\
$^{2}$Department of Astronomy, University of California at Berkeley, Berkeley, CA 94720, USA\\
$^{3}$Flatiron Institute, Simons Foundation, 162 Fifth Ave, New York, NY 10010, USA\\
$^{4}$Department of Earth Sciences, University of Hawai`i at M\={a}noa, Honolulu, HI, USA\\
$^{5}$NASA Ames Research Center, Moffett Blvd, Mountain View, CA 94035, USA\\
$^{6}$Institute of Astronomy, University of Cambridge, Madingley Road, CB3 0HA,  Cambridge, UK\\
$^{7}$Department of Astronomy, The University of Texas at Austin, Austin, TX 78712, USA\\
$^{8}$Department of Physics, University of Warwick, Gibbet Hill Road, Coventry, CV4 7AL, UK\\
$^{9}$Centre for Exoplanets and Habitability, University of Warwick, Gibbet Hill Road, Coventry, CV4 7AL, UK\\
$^{10}$Institute for Astronomy, University of Hawai`i at M\={a}noa, Honolulu, HI, USA\\
$^{11}$Department of Physics and Astronomy, University of North Carolina at Chapel Hill, Chapel Hill, NC 27599, USA\\
$^{12}$Department of Physics and Astronomy, University of California, Los Angeles, Los Angeles, CA 90095, USA\\
$^{13}$Universit\'e Grenoble-Alpes, CNRS Institut de Plan\'etologie et d'Astrophysique (IPAG), F-38000 Grenoble, France\\
$^{14}$Jet Propulsion Laboratory, California Institute of Technology 4800 Oak Grove Dr., Pasadena, CA 91109, USA\\
$^{15}$University of Rochester, Department of Physics and Astronomy, Rochester, NY 14627, USA\\
$^{16}$Joint ALMA Observatory, Avenida Alonso de C\'ordova 3107, Vitacura, Santiago, Chile\\
$^{17}$Steward Observatory, University of Arizona, Tucson, AZ, 85719, USA\\
}

\pubyear{2019}

\begin{document}
\label{firstpage}
\pagerange{\pageref{firstpage}--\pageref{lastpage}}
\maketitle

\begin{abstract}
Dippers are a common class of young variable star exhibiting day-long dimmings with depths of up to several tens of percent. A standard explanation is that dippers host nearly edge-on ($i_{\rm d}\approx70$\degrees{}) protoplanetary discs that allow close-in ($<$1~au) dust lifted slightly out of the midplane to partially occult the star. The identification of a face-on dipper disc and growing evidence of inner disc misalignments brings this scenario into question. Thus we uniformly (re)derive the inclinations of 24 dipper discs resolved with (sub-)mm interferometry from ALMA. We find that dipper disc inclinations are consistent with an isotropic distribution over $i_{\rm d}\approx0-75$\degrees{}, above which the occurrence rate declines (likely an observational selection effect due to optically thick disc midplanes blocking their host stars). These findings indicate that the dipper phenomenon is unrelated to the outer ($>$10~au) disc resolved by ALMA and that inner disc misalignments may be common during the protoplanetary phase. More than one mechanism may contribute to the dipper phenomenon, including accretion-driven warps and ``broken" discs caused by inclined (sub-)stellar or planetary companions.
\end{abstract}



\section{Introduction} 
\label{sec:intro}

Photometric variability is a hallmark of young ($\lesssim10$~Myr) stars and studies of this variability provide insight into the physical processes underpinning early stellar evolution and planet formation. Some young stars transiently dim due to obscuration by circumstellar dust, a phenomenon first noted in ground-based photometry of bright intermediate-mass Herbig Ae/Be stars that fade up to several orders of magnitude for weeks to months, named UXOR variables after the archetype UX Orionis \citep{Herbst1994}. More sensitive space-based telescopes---most notably {\it CoRoT} \citep{Baglin2006}, {\it Spitzer} \citep{Fazio2004}, and {\it Kepler} \citep{Borucki2010}---later identified a related class of fainter, typically late-type pre-main sequence stars known as "dippers" \citep{Alencar2010,MC2011,Cody2014,Ansdell2016a,Stauffer2017}. The dippers exhibit more moderate dimming events, with depths up to several tens of percent and durations of roughly one day, and can be either quasi-periodic or episodic \citep[e.g.,][]{Cody2014} with diverse profile shapes \citep[e.g., see Figure 4 in][]{Ansdell2016a}. Dippers are particularly interesting for studying disc evolution and planet formation as they are common, making up 20--30\% of young stellar populations \citep{Alencar2010, Cody2014}.

The dipper phenomenon is thought to result from partial occultations of the star by circumstellar dust. The dust interpretation is supported by dippers invariably having infrared emission in excess of that expected from the stellar photosphere, indicating the presence of a protoplanetary disc, as well as the dips often being shallower at longer wavelengths, where dust is less scattering \citep{MC2011,Cody2014,Schneider2018}. An origin in the inner ($<$1~au) disc is suggested by the quasi-periodic dippers having periods of a few days, which is often indistinguishable from the stellar rotation period \citep{Bodman2017}, as well as a positive correlation between dip depth and excess emission in the \emph{WISE} 4.6$\mu$m band, which is sensitive to warm dust grains near the disc-star co-rotation radius around late-type stars \citep{Ansdell2016a}. 

The physical mechanism identified early on to explain the prototypical dipper AA~Tau invokes magnetospheric accretion to lift dusty material out of the disc midplane, creating an inner warp that occults the star \citep{Bouvier1999}. A prediction of this model is that discs around dipper stars should be observed at nearly edge-on ($i_{\rm d}\approx70$\degrees{}) inclinations, since lower inclinations would preclude the occultation while higher inclinations would cause the optically thick disc midplane to obscure the star entirely. Indeed, \cite{McGinnis2015} and \cite{Kesseli2016} reproduced the light curves of dippers in NGC~2264 obtained by {\it Spitzer} and {\it CoRoT} using models of magnetospheric accretion from nearly edge-on protoplanetary discs. \cite{Bodman2017} later revised this accretion warp model with magnetospheric truncation theory to show that it could explain dippers with discs of only moderate inclinations down to $i_{\rm d}\approx50$\degrees{}.

The expectation of dipper systems tending to have high inclinations can be tested by resolving the discs with infrared or ${\rm (sub-)mm}$ interferometry. Although the infrared probes closer to the star ($<$1~au) where the dipper phenomenon likely originates, the faintness of the dippers often prohibits these observations. This is not the case for (sub-)mm interferometry, which with the advent of the Atacama Large sub-millimeter/Millimeter Array (ALMA) can now quickly resolve protoplanetary discs around all stellar types, but is sensitive to the outer ($>$10~au) disc. In our previous work \citep{Ansdell2016b}, we used archival ALMA data of three dippers to show that their outer discs ranged from face-on to edge-on. This hinted toward significantly misaligned inner disc components and/or the need for other dipper mechanisms. Evidence of misaligned inner discs has also been recently inferred for several systems using high-contrast optical/infrared images, which have revealed shadows in the outer disc cast by unseen inclined inner disc components \citep[e.g.,][]{Marino2015,Stolker2016,Debes2017,Benisty2018,Casassus2018}. A notable example is the dipper J1604, which hosts a face-on transition disc resolved by ALMA \citep{Ansdell2016b} and variable shadows seen by VLT/SPHERE \citep{Pinilla2018a}, suggesting a highly misaligned ($\sim$70--90\degrees{}) and dynamic inner disc component.

In this work, we uniformly (re)analyze resolved (sub-)mm ALMA data for two dozen dipper discs in an effort to robustly infer the distribution of their outer disc inclinations. In Section~\ref{sec:sampledata} we present our sample and describe the datasets used in this work. We derive disc inclinations from the ALMA data in Section~\ref{sec:analysis}, then in Section~\ref{sec:discussion} we construct the dipper disc inclination distribution and discuss its impact on our understanding of the dippers and inner disc misalignments. We conclude in Section~\ref{sec:summary} and suggest avenues for future work.


\section{Sample \& Datasets} 
\label{sec:sampledata}

\subsection{Sample} 
\label{sec:sample}

Our sample consists of all known dippers in the $\rho$~Ophiuchi ($\rho$~Oph) and Upper Scorpius (Upper Sco) star-forming regions that have been identified by their {\it K2} Campaign 2 ({\it K2}/C2) light curves and have discs resolved by ALMA. We focus on these two nearby \citep[$\approx$130~pc;][]{Gagne2018} star-forming regions because they are known to host numerous dippers \citep[e.g.,][]{Cody2018} and have been surveyed extensively with ALMA \citep[e.g.,][]{Barenfeld2016, Cieza2018}. Moreover, they were observed during the same {\it K2} campaign and thus contain the same systematics in their light curves (see Section~\ref{sec:light curves}).

The dippers in our sample were all previously identified based on their {\it K2}/C2 light curves by \cite{Ansdell2016a}, \cite{Hedges2018}, and/or \cite{Cody2018} using different methods. \cite{Ansdell2016a} worked with citizen scientists to identify 25 dippers by eye and study 10 of them in detail. \cite{Hedges2018} then used these 25 dippers to train a supervised machine learning algorithm with a random forest classifier to expand the sample to 95 dippers. \cite{Cody2018} employed their traditional statistics of periodicity and symmetry to categorize variable young stars in the {\it K2}/C2 dataset, identifying 94 dippers. Combining these samples results in 122 unique dippers.

\afterpage{
\begin{landscape}
\begin{table}
\small
\caption{Dippers with Resolved Discs}
\begin{threeparttable}
\renewcommand*{\arraystretch}{1.4} 
\label{tab:results}
\begin{tabular}{lccccccccrrc}
\hline \hline
\multicolumn{1}{c}{EPIC} &
\multicolumn{1}{c}{2MASS} &
\multicolumn{1}{c}{Name} &
\multicolumn{1}{c}{Region} &
\multicolumn{1}{c}{SpT} &
\multicolumn{1}{c}{Ref. SpT$^{a}$} &
\multicolumn{1}{c}{Ref. Dipper$^{a}$} &
\multicolumn{1}{c}{$i_{\rm d, lit}^{b}$} &
\multicolumn{1}{c}{Ref. $i_{\rm d, lit}$} &
\multicolumn{1}{c}{$i_{\rm d}^{b}$} &
\multicolumn{1}{c}{${\rm P.A.}_{\rm d}^{c}$} &
\multicolumn{1}{c}{ALMA ID} \\
\hline
203937317 & J16261706-2420216 & DoAr 24 & Oph & K7.5 & A16 & A16, HHK18, CH18 & ... & ... & $7^{+6}_{-5}$ & ... & 2016.1.00336.S \\ 
204638512 & J16042165-2130284 & J1604 & Usc & K2   & L12 & A16, HHK18, CH18 & 6.0$\pm$1.5 & M12  & $7.8^{+0.1}_{-0.1}$ & $-6^{+1}_{-1}$ & 2017.1.01180.S  \\ 
204281213$^c$ & J15583692-2257153 & HD 143006 & Usc & G5IVe & P16 & CH18 & 18.6$\pm$0.8 & H18 & $19.4^{+1.0}_{-1.1}$ & ... & 2013.1.00395.S \\ 
203770559 & J16250208-2459323 & WSB 19 & Oph & M4.5 & E11 & HHK18 & 34.0$\pm$8.2  & C18 & $26.7^{+3.9}_{-4.5}$ & $-5^{+8}_{-8}$ & 2016.1.00545.S \\
203770559$^d$ & J16250208-2459323 & WSB 19B & ... & ...  & ... & ...   & 33.2$\pm$14.9 & C18 & $43.7^{+4.2}_{-4.5}$ & $-126^{+6}_{-5}$ & 2016.1.00545.S \\
205345560 & J16062383-1807183 & ... & Usc & ...  & ... & CH18 & ... & ...  & $28.9^{+9.8}_{-9.3}$ & $93^{+23}_{-22}$ & 2018.1.00564.S \\
204630363 & J16100501-2132318 & EM* StHA 123 & Usc & K7.5 & R15 & A16 & ... & ... & $38.0^{+0.1}_{-0.1}$ & $60^{+1}_{-1}$ & 2016.1.00336.S \\
204864076 & J16035767-2031055 & RX J1603.9-2031A & Usc & K5 & L12 & A16, HHK18, CH18 & $69^{+21}_{-27}$ & B17 & $45.4^{+6.0}_{-7.0}$ & $41^{+8}_{-7}$ & 2016.1.00336.S \\
204176565 & J16221852-2321480 & V$^*$ V935 Sco & Oph & K5 & C10 & HHK18, CH18 & 43.8$\pm$3.1 & C18 & $47.2^{+1.0}_{-1.0}$ & $84^{+1}_{-1}$ & 2016.1.00545.S \\ 
203936815 & J16264285-2420299 & ISO-Oph 62 &  Oph & M1  & E11 & HHK18, CH18 & 59.9$\pm$11.9 & C18 & $47.4^{+5.2}_{-5.5}$ & $154^{+6}_{-7}$ & 2016.1.00545.S \\ 
203936815$^d$ & J16264285-2420299 & ISO-Oph 62B & ... & ... & ... & ... & ... & ... & $66.9^{+15.6}_{-20.0}$ & $145^{+20}_{-23}$ & 2016.1.00545.S \\
203950167 & J16230923-2417047 & IRAS 16201-2410 & Oph & G0 & M10 & HHK18, CH18 & 51.6$\pm$4.7 & C17 & $48.4^{+0.5}_{-0.3}$ & $80^{+1}_{-1}$ & 2016.1.00545.S \\
204142243 & J16222497-2329553 & WSB 14 & Oph & ... & ... & HHK18, CH18 & ... & ...  & $48.4^{+4.4}_{-4.5}$ & $162^{+5}_{-5}$ & 2016.1.00545.S \\ 
203962599 & J16265677-2413515 & ISO-Oph 83 & Oph & K7 & E11 & HHK18, CH18 & 63.6$\pm$2.8 & C18 & $51.6^{+2.6}_{-2.8}$ & $169^{+3}_{-3}$ & 2016.1.00545.S \\ 
205151387 & J16090075-1908526 & UCAC2 24371748 & Usc & M1.0 & A16 & A16, HHK18, CH18 & $56^{+5}_{-5}$ & B17 & $54.4^{+6.5}_{-7.7}$ & $154^{+8}_{-8}$ & 2011.0.00526.S \\ 
203860592 & J16273942-2439155 & WSB 52 & Oph & K5 & W05 & CH18 & 54.3$\pm$0.3 & H18 & $53.9^{+0.4}_{-0.4}$ & $142^{+1}_{-1}$ & 2016.1.00545.S  \\ 
205238942 & J16064794-1841437 & ... & Usc & M0.0 & R15 & HHK18, CH18 & ... & ... & $55.5^{+0.1}_{-0.1}$ & $20^{+1}_{-1}$ & 2018.1.00564.S \\
204489514 & J16030161-2207523 & ... & Usc & M2.7 & M17 & A16, HHK18, CH18 & $52^{+22}_{-42}$ & B17 & $59.3^{+8.4}_{-9.2}$ & $27^{+7}_{-7}$ & 2016.1.00336.S \\ 
204514548$^c$ & J15564002-2201400 & HD 142666 & Usc & A8 & F15 & CH18 & 62.2$\pm$0.1 & H18 & $61.2^{+0.5}_{-0.5}$ & $161^{+1}_{-1}$ & 2013.1.00498.S \\ 
203895983 & J16041893-2430392 & [M81] I-490 & Usc & M2.5  & R15 & A16, HHK18 & ... & ...  & $62.6^{+3.1}_{-4.0}$ & $67^{+4}_{-4}$ & 2018.1.00564.S \\
203843911 & J16262367-2443138 & DoAr 25 & Oph & K5 & W05 & A16, HHK18, CH18 & 67.4$\pm$0.2 & H18 & $66.3^{+0.1}_{-0.1}$ & $111^{+1}_{-1}$ & 2016.1.00336.S \\
203824153 & J16285407-2447442 & WSB 63  & Oph & M1.5 & E11 & A16, HHK18, CH18 & 66.3$\pm$1.5 & C17 & $67.3^{+0.5}_{-0.5}$ & $1^{+1}_{-1}$ & 2016.1.00336.S \\ 
204399980 & J16131158-2229066 & HD 145718 & Usc & A5 & L12 & CH18 & ... & ... & $70.4^{+1.2}_{-1.2}$ & $1^{+1}_{-1}$ & 2015.1.01600.S \\ 
204211116 & J16214199-2313432 & ... & Oph & M3 & V16 & A16, HHK18, CH18 & ... & ... & $71.3^{+3.0}_{-3.0}$ & $38^{+2}_{-2}$ & 2016.1.00336.S \\
205080616 & J16082324-1930009 & UCAC2 24134752 & Usc & K9 & P01 & CH18 & $74^{+5}_{-4}$ & B17  & $70.6^{+5.8}_{-5.7}$ & $124^{+3}_{-4}$ & 2011.0.00526.S \\
203850058 & J16270659-2441488 & ISO-Oph 102 & Oph & M5 & M15 & HHK18, CH18 & 73$\pm$23 & A16b & $84.0^{+4.1}_{-3.9}$ & $13^{+1}_{-1}$ & 2012.1.00046.S \\ 
\hline
\end{tabular}
\begin{tablenotes}
\scriptsize
\item $^{a}$ References. A16$=$\cite{Ansdell2016a}; A16b$=$\cite{Ansdell2016b}; B17$=$\cite{Barenfeld2017}; CH18$=$\cite{Cody2018}; C18$=$\cite{Cieza2018}; C17$=$\cite{Cox2017}; C10$=$\cite{Cieza2010}; E11$=$\cite{Erickson2011}; F15$=$\cite{Fairlamb2015}; H18$=$\cite{Huang2018}; HHK18$=$\cite{Hedges2018}; L12$=$\cite{LM2012}; M17$=$\cite{Martinez2017} M15$=$\cite{Manara2015}; M12$=$\cite{Mathews2012}; P16$=$\cite{Pecaut2016}; P01$=$\cite{Preibisch2002}; R15$=$\cite{Rizzuto2015}; V16$=$\cite{vdP2016}; W05$=$\cite{Wilking2005}.
\item $^{b}$ Disc inclinations in degrees: $i_{\rm d, lit}$ are disc inclinations derived from ALMA data in the literature (references given in the Ref. $i_{\rm d, lit}$ column); $i_{\rm d}$ are the disc inclinations uniformly (re-)derived in this work using {\tt GALARIO} to fit the ALMA datasets (Project IDs given in the ALMA ID column). The disc inclinations from C18 and C17 are calculated from the measured semi-major and semi-minor axes reported in those works, while those from H18 are their 2D Gaussian fits.
\item $^{c}$ Position angle in degrees, measured positive East of North. We do not provide P.A. values for EPIC~203937317 or EPIC~204281213; the former was unconstrained by our fit due to its face-on geometry and the latter was unreliable due to the disc asymmetry. 
\item $^{d}$ EPIC~204281213 and 204514548 are listed under alternative {\it K2} IDs of EPIC~204281210 and 204514546, respectively, in \cite{Cody2018}. We use the former as they are associated with {\it K2} C15 light curves shown in Figure~\ref{fig:data}.
\item $^{d}$ These are the secondary components of the two binary discs resolved in the ALMA data (Section~\ref{sec:discinc}).
\end{tablenotes}
\end{threeparttable}
\end{table}
\end{landscape}
}

The issue with combining these samples is that significantly different criteria were used to identify the dippers, and sometimes the methods did not all agree. Therefore we consider all the dippers in the combined samples, then require at least three dips to be present in the {\it K2}/C2 light curves and apply the $R_{\rm dip} > 5.0$ selection criteria of \cite{Ansdell2016a}. To summarize the more detailed description of $R_{\rm dip}$ in \cite{Ansdell2016a}: $R_{\rm dip}$ is the ratio of the average of the three deepest dips to the standard deviation for a normalized light curve that has been put through a high-pass filter with a cut-on frequency of 1 day$^{-1}$ (the high-pass filter preserves the dips while suppressing the periodic variability from stellar rotation due to the different duty cycles). This criteria helps to avoid stars with large intrinsic variability (e.g., stochastic variables) as well as noisy light curves with suspicious irregularities related to instrumental effects (e.g., charge bleed) or data corruption (e.g., data discontinuities). This cut reduces the sample to 79 dippers.  

We then use data from our targeted ALMA programme as well as a search of the ALMA archive to identify those with resolved circumstellar discs (see Section~\ref{sec:alma} for a description of the ALMA datasets used in this work, Section~\ref{sec:discinc} for the derivation of the disc inclinations, and Appendix~\ref{appendix-C} for a discussion of the dippers observed but not resolved by ALMA). We note that a general literature search did not return any dippers with disc inclinations derived only from pre-ALMA radio interferometers, and only one known dipper has an inclination derived only from scattered light \cite[EPIC~204206295 or DoAr~28 with a modestly inclined disc of $i_{\rm d}=50^{+15}_{-10}$;][]{Rich2015}, which we do not include in our sample. Our final sample therefore consists of 24 dippers, 12 in $\rho$ Oph and 12 in Upper Sco. Their basic properties are presented in Table~\ref{tab:results} and their ALMA data and {\it K2}/C2 light curves are shown in Figure~\ref{fig:data}.

\subsection{K2 Light Curves} 
\label{sec:light curves}

To construct the {\it K2}/C2 light curves used in this work, we re-extract the photometry for each dipper from the original pixel-level data. This is needed because {\it K2} \citep{Howell2014}, the successor mission to {\it Kepler} after the spacecraft lost two out of its four reaction wheels, adopted an ecliptic-observing orientation to stabilize its pointing using solar radiation pressure as a pseudo third reaction wheel. Due to Sun-angle constraints, {\it K2} observations were organized into a series of sequential observing campaigns, which were limited to fields located around the ecliptic plane and to durations of roughly 80 days. Quasi-periodic thruster firings throughout each observing campaign were then needed to correct for residual pointing drift, known as ``roll motion," which caused characteristic ``sawtooth" patterns in the simple aperture photometry (SAP) light curves as targets moved around in their fixed apertures. 

To correct for these effects, we use a modified Pixel Level Decorrelation (PLD) method to remove the roll motion noise while preserving intrinsic, astrophysical variability. First, we use the \texttt{interact} tool in the {\tt LIGHTKURVE} package \citep{LightKurve2018} to hand-select pixel apertures, which are customized to include as much of the target flux as possible while avoiding nearby contaminants. We then sum the flux within these custom apertures, which are typically a few {\it Kepler} pixels across (where one {\it Kepler} pixel is 4\as{}$\times$4\as{}) to build the SAP light curves. Decorrelation matrices are then generated from: 1) pixel time-series of neighboring, quiet targets, which by nature strongly exhibit the {\it K2} roll motion pattern; 2) fourth-order polynomials in time to capture extremely long term ($>$50-day) variability due to changes in spacecraft temperature and velocity aberration; and 3) two-dimensional, fourth-order polynomials of the point-spread function centroids in column and row (measured using {\tt LIGHTKURVE}). We apply these decorrelation matrices in the same way as the PLD method from \cite{Luger2016}, where optimum weights are derived for each component of the matrix using linear algebra. We split each light curve in half (at cadence number 97682) and fit the weights separately to each half; this accounts for a shift in the light curve noise properties, which is commonly seen in the {\it K2} data due to the change in Sun-angle on the spacecraft (and thus roll motion direction) at approximately the centre of each campaign. Along with the weights, we also simultaneously fit a Gaussian Process to the light curve in order to capture the astrophysical variability of the dippers, which is frequently orders of magnitude greater than the spacecraft systematics. The best-fit decorrelation matrices are then weighted and summed to build our spacecraft motion correction. Using this method, any short term variability due to spacecraft motion is removed, while astrophysical variability is preserved.

We note that for the two bright A-type stars in our sample (EPIC~204514548 and 204399980) the above method could not be applied due to saturation and flux bleed issues. Instead we use the publicly available {\it K2} Self Field Flattening (SFF) light curves described in \cite{Vanderburg2014} and made available through the Mikulski Archive for Space Telescopes (\url{http://archive.stsci.edu/kepler}), which should be sufficient as the spacecraft motion noise is much smaller than the astrophysical trends for these bright stars. We also note that four dippers (EPIC~203770559, 203895983, 203936815, and 203950167) could not be separated from their bright, nearby companions (see Table~\ref{tab:ao}) using aperture photometry given the large size of the {\it Kepler} pixels, although in some cases we detect both disc components around the primary and secondary; we discuss these candidate binary systems in Section~\ref{sec:mult}.

\begin{figure*}
\begin{center}
\includegraphics[width=16cm]{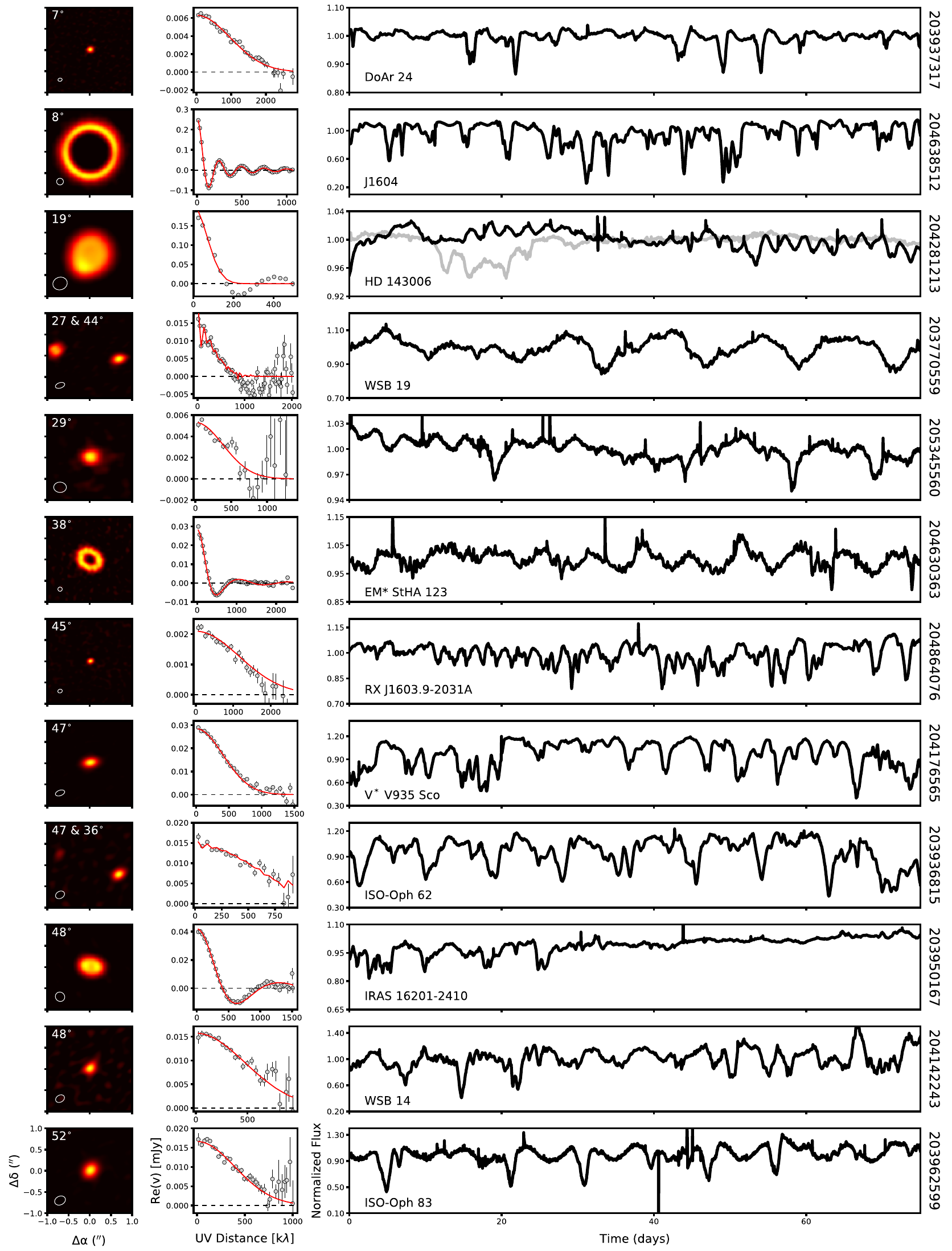}
\caption{ALMA images (left) and associated visibilities (middle) as well as {\it K2}/C2 light curves (right) for the sources in our sample. The ALMA images are $2^{\prime\prime}\times2^{\prime\prime}$ (corresponding to about 260 $\times$ 260 au) with the beam shown by the white ellipse and the disc inclination given for reference. The {\tt GALARIO} model fits are shown by the red line over the visibility data, which are de-projected using the best-fit {\tt GALARIO} geometries; declining visibilities with UV distance indicate resolved sources (for the binaries we show the combined visibilities). The EPIC names of each source are on the far right, and common names are given in the {\it K2}/C2 panels (see Table~\ref{tab:results}). The {\it K2}/C15 light curves for EPIC~204281213, 204489514, and 204514548 are overlaid over their {\it K2}/C2 data.}
\label{fig:data}
\end{center}
\end{figure*}

Three dippers in our sample (EPIC~204281213, 204489514, and 204514548) also had a second epoch of {\it K2} data taken during Campaign 15 ({\it K2}/C15), which was conducted three years after {\it K2}/C2. We plot the {\it K2}/C15 SFF light curves for these sources over their {\it K2}/C2 data in Figure~\ref{fig:data}, illustrating how dipper behavior can change on the timescale of years. All three sources would still be classified as dippers based on the selection criteria applied in Section~\ref{sec:sample} when using their {\it K2}/C15 light curves.

\renewcommand{\thefigure}{\arabic{figure} (Cont.)}
\addtocounter{figure}{-1}
\begin{figure*}
\begin{center}
\includegraphics[width=16.5cm]{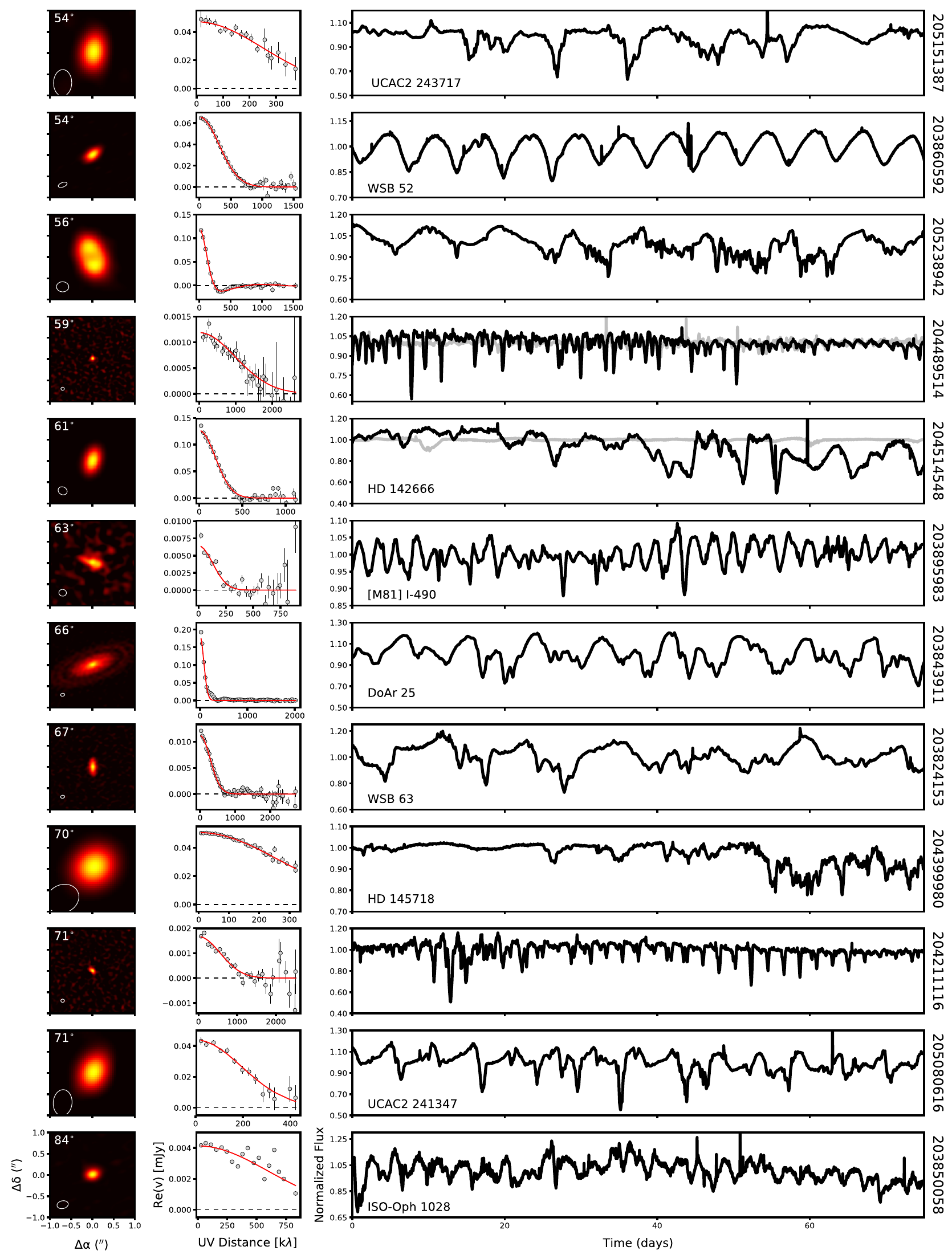}
\caption{}
\end{center}
\end{figure*}
\renewcommand{\thefigure}{\arabic{figure}}

\subsection{Sub-mm/mm ALMA Data} 
\label{sec:alma}

The ALMA data used to resolve the dipper discs in our sample come from both our targeted ALMA programme as well as archival ALMA programmes, as described below. Data calibration and imaging are performed using CASA; the data were pipeline calibrated by NRAO staff and include standard flux, phase, bandpass, and gain calibrations. The ALMA Project IDs of the data used for each dipper are given in Table~\ref{tab:results}.

Our targeted ALMA programme to resolve dipper discs (Project ID: 2016.1.00336.S; PI: Ansdell) was a high-resolution Band~6 ($\lambda\approx1.3$~mm) survey of nine dippers conducted in Cycle~4 using the C40-7 configuration (21--3638~m baselines). The continuum spectral windows were centred on 233.29, 220.40, and 217.47~GHz with bandwidths of 2.00, 2.00, and 1.88~GHz, respectively. The programme was split into two Science Goals, one for the four brighter ($F_{\rm 1.3mm}\gtrsim10$~mJy) dippers (EPIC~204630363, 203937317, 203843911, and 203824153) and another for the five fainter ($F_{\rm 1.3mm}\approx1$~mJy) dippers (EPIC~204107757, 204489514, 204864076, 204757338, and 204211116) in the sample. The four brighter targets were observed on 15 August 2017 with 45 12-m antennas and 4.7~min on-source integration times for a mean continuum rms of 0.07 mJy beam$^{-1}$. The five fainter targets were observed on 18 Aug 2017 with 42 12-m antennas and 8.5~min on-source integration times for a mean continuum rms of 0.04 mJy beam$^{-1}$. The 0.1\as{} ($\sim$10~au) angular resolution of our observations were sufficient to resolve all the dipper discs targeted by our programme, except for those around EPIC~204107757 and 204757338, which were therefore not included in our sample.

We also make significant use of archival ALMA data, in particular those taken for the previously published large-scale surveys of Upper Sco \citep{Carpenter2014,Barenfeld2016, Barenfeld2017} and $\rho$~Oph \citep{Cieza2018,Williams2019}. The $\rho$~Oph survey (Project ID: 2016.1.00545.S; PI: Cieza) was conducted in ALMA Band~6 ($\lambda\approx1.3$~mm) during Cycle~4 and the sample was split into two Science Goals: the brighter and less evolved sources were observed at higher resolution (0.25\as{}) and sensitivity (0.15~mJy~beam$^{-1}$ continuum rms), while the fainter and more evolved sources were observed at lower resolution (0.8\as{}) and sensitivity (0.25~mJy~beam$^{-1}$ continuum rms). The Upper Sco surveys were conducted in ALMA Band~7 ($\lambda\approx880~\mu$m) in Cycle~0 and Cycle~2 (Project IDs: 2011.0.00966.S, 2013.1.00395.S). The observations had angular resolutions between 0.35\as{} and 0.73\as{} with a median of 0.37\as{}, and continuum rms values ranging from 0.13~mJy~beam$^{-1}$ to 0.26~mJy~beam$^{-1}$ with a median of 0.15~mJy~beam$^{-1}$. Additionally, we make use of data taken for a new ALMA survey conducted to complete the (sub-)mm census of Upper Sco as new disc-hosting members have been discovered (Project ID: 2018.1.0056.S; PI: Carpenter); these data were taken in ALMA Band~7 during Cycle~6 with typical angular resolutions of 0.2\as{} ($\sim$25~au) and continuum rms values of 0.15~mJy~beam$^{-1}$.

We also use data from selected archival ALMA programmes when they offered higher spatial resolution. Thus the data for EPIC~203850058, 204638512, and 204514548 come from ALMA programmes 2012.1.00046.S, 2017.1.01180.S, and 2013.1.00498.S with PIs Phan-Bao, Loomis, and Perez, respectively. Finally, although we do not perform model fits to these data in this work, four dippers in our sample (EPIC~203843911 or DoAr~25, EPIC~203860592 or WSB~52, EPIC~204281213 or HD~143006, and EPIC~204514548 or HD~142666) have higher resolution ($\sim$5~au) ALMA observations taken as part of the Disc Substructures at High Angular Resolution Project \citep[DSHARP;][]{Andrews2018}. In Section~\ref{sec:dsharp}, we discuss these DSHARP data within the context of this work.

\begin{figure*}
\begin{center}
\includegraphics[width=17.5cm]{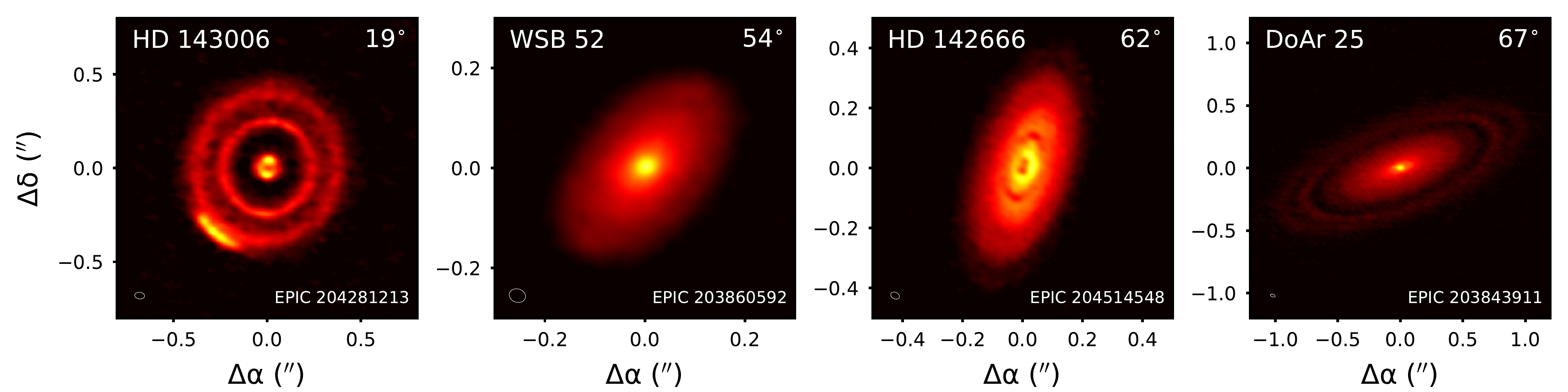}
\caption{High-resolution ($\sim$5~au) ALMA images of the four dippers in our sample included in DSHARP. Names and inclinations from DSHARP \citep{Huang2018} are given at the top of each panel, EPIC numbers are shown in the lower right, and the beam size is illustrated by the white ellipse in the lower left.}
\label{fig:dsharp}
\end{center}
\end{figure*}

\subsection{Adaptive Optics Imaging} 
\label{sec:ao}

All but two of the dippers in our sample (EPIC 204142243 and 205345560) have been inspected for close companions with high-contrast imaging (see Table~\ref{tab:ao} and Table~\ref{tab:contrasts}). Ten of these were observed and/or analyzed as part of this work using adaptive optics (AO) imaging with the Near-infraRed imaging Camera (NIRC2) mounted on the 10-m Keck II telescope atop Maunakea. For the sources we observed with NIRC2, those with $R<13.5$ used natural guide star AO \citep{Wizinowich2000,vanDam2004}, while the fainter sources used laser guide star AO \citep{Wizinowich2006,vanDam2006}. Imaging was done with the narrow camera and several sources also used non-redundant aperture masking (NRM).

For the NIRC2 data reduction, each frame is linearized and corrected for geometric distortion using the solution from \cite{Yelda2010}, then the four image quadrants are used to de-bias the ``stripe noise" (i.e., spatially correlated readnoise) that results from electronics noise during readout being mirrored in each quadrant. Images are then dark-subtracted and flat-fielded using the most contemporaneous available calibration files, and  ``dead" and ``hot" pixels are interpolated over. Dead pixels are identified from ``super-flats" taken in 2006--2013 as any pixel with a response $<$30~percent in at least half of all super-flats. Similarly, hot pixels are identified from a comparable set of ``super-darks" as any pixel with $\ge$10 counts in at least half of the super-darks. Pixels with flux levels $>$10$\sigma$ above the median of the 8 adjacent pixels are flagged as cosmic rays or transient hot pixels and replaced with the median.

Our analysis of the standard AO imaging data broadly follows the methods of \cite{Kraus2016}. To detect faint and wide ($\ga$500 mas) companions, we subtract an azimuthal median point spread function (PSF) model. This adds no additional noise at wide separations, but leaves speckles in places, making it non-ideal for detecting close-in companions. To probe smaller inner working angles, we construct and subtract the best-fitting empirical PSF of another (single-star) target taken from among the 1000 images in the same filter nearest in time that were publicly available in the Keck Observatory Archive. In each individual frame, we measure the flux within 40 mas radius apertures centred on every image pixel, and compute the corresponding noise statistics as a function of radius by measuring the standard deviation of those fluxes in five-pixel-wide annuli around the primary. We then stack the detection-significance maps with a weighted mean, flagging any pixel with $>$6$\sigma$ confidence as the location of a candidate companion. 

Those candidates are then visually inspected to reject erroneous detections due to remaining cosmic rays and hot pixels as well as imperfect PSF subtraction of the strongest super-speckles. If any genuine companions are located within the speckle pattern of their primary star, the empirical PSF routine is repeated with a binary model that iteratively fits for the separation, position angle, and contrast of the two sources, then tests the reference PSFs (doubled using that binary model) to find the best-fitting empirical PSF template, repeated until the same best-fitting PSF template produces the same best-fitting binary model.

The analysis of the NRM observations broadly follows the methods of \cite{Kraus2008} and \cite{Ireland2013}. The NRM observations use a pupil plane mask to resample the telescope into a sparse interferometric array. This allows the use of the complex triple product, or closure-phase observable, to remove non-common path errors produced by atmospheric conditions or variable optical aberrations. To remove systematics in the closure-phase observable, observations of the science targets are paired with calibration observations of other stars nearby in time, which were typically also science targets. Binary system profiles are then fit to the calibrated closure phases to measure component separations and position angles and calculate contrast limits.

The remaining dippers with high-contrast imaging were observed as part of previous surveys (see Table~\ref{tab:ao} and Table~\ref{tab:contrasts}). We defer to those works for details of the data reduction and analysis and adopt their reported results in this work.


\section{Analysis} 
\label{sec:analysis}

\subsection{Outer Disc Inclinations}
\label{sec:discinc}

Although many sources in our sample have disc inclinations derived from ALMA data reported in the literature (see Table~\ref{tab:results}), these were derived with disparate methods, and there are some inconsistencies among reported values. For example, the disc inclinations for $\rho$~Oph targets from \cite{Cieza2018} and \cite{Cox2017} must be inferred from the reported semi-major and semi-minor axes of 2D Gaussian model fits to the observed visibilities using CASA's {\tt uvmodelfit} task. The disc inclinations of Upper Sco members from \cite{Barenfeld2017}, on the other hand, use a Markov chain Monte Carlo (MCMC) approach to fit the observed visibilities to synthetic visibilities derived from a self-consistent disc model with an assumed dust surface density profile parameterized by a truncated power-law. The latter method is more physically motivated, but can be limiting for moderate signal-to-noise and/or marginally resolved discs, which are common in large-scale ALMA disc surveys. However, the former method is not appropriate for discs with large inner cavities (there are a few of these ``transition discs" in our sample; see Figure~\ref{fig:data}) and also does not provide posterior distributions that are useful for our analysis of the disc inclination distribution (see Section~\ref{sec:incdist}).

Thus we (re)derive the disc inclinations for our entire sample using the {\tt GALARIO} computational library \citep{Tazzari2018} combined with the {\tt emcee} package \citep{emcee}, which allows us to quickly fit 2D disc models to the ALMA visibilities by enabling the rapid exploration of parameter space. {\tt GALARIO} is a Python library that uses GPUs, or alternatively multiple CPU cores, to speed up the computation of synthetic visibilities. Because of its modular structure, we can use {\tt GALARIO} for the likelihood computation in {\tt emcee}, a Python implementation of MCMC Ensemble sampling for Bayesian parameter estimation. We use the GPU version of {\tt GALARIO}, which is $\sim$150$\times$ faster than standard Python implementations that rely on {\tt scipy} and {\tt numpy} packages. \cite{Tazzari2018} provides a detailed explanation of employing {\tt GALARIO} to fit interferometric visibilities like those from ALMA and we follow their general implementation procedure. The fits typically use 120 walkers and 5000 steps for the MCMC; the mean accepted fraction of steps are all between 0.2 and 0.5, implying that the chains are converging.

Most of the dippers in our sample are single stars whose discs lack resolved structure (e.g., no gaps or rings), thus we most often fit the visibilities using a simple 2D Gaussian disc model with six free parameters: a flux normalization term, the full-width-half-max along the semi-major axis, the inclination ($i_{\rm d}$) and position angle (P.A.$_{\rm d}$) of the disc, and the offset in right ascension and declination of the source centre from the phase centre. For the two resolved binary discs in our sample (EPIC 203770559 and 203936815), we simultaneously fit two 2D Gaussians to the data. Four dippers (EPIC 203950167, 204630363, 204638512, and 205238942) exhibit large inner cavities; for these, we use the aforementioned 2D Gaussian model, but with one more free parameter, an inner cutoff radius. EPIC~203843911 and EPIC~204281213 exhibit more complex features in the data that our simple models could not account for (i.e., narrow gaps/rings and azimuthal asymmetries, respectively; see Figure~\ref{fig:data}). However, in both cases our inclinations agree well with those derived by \cite{Huang2018} from the extremely high resolution DSHARP observations of these objects. 

Indeed, for all four dippers in our sample that overlap with DSHARP (see Figure~\ref{fig:dsharp}), our derived disc inclinations match to within $\sim$1\degrees{} of those from \cite{Huang2018}. This instills general confidence in the precision of our inclinations, which are derived from simple models of more moderate resolution data. This is consistent with the general findings of \cite{Huang2018}, who performed both simple 2D elliptical Gaussian fits to the DSHARP discs as well as more detailed fits to the individual well-resolved annular substructures within each disc, finding that the derived inclinations agreed to within roughly a degree in all cases, with no apparent biases in the results. This suggests that any effects from fitting the simple 2D Gaussian models in this work (e.g., due to unaccounted model-dependent uncertainties) is on the level of $\lesssim$1\degrees{} and thus should not impact our analysis, which focuses on population-level statistics.

\begin{table*}
\centering
\caption{Candidate Companions Detected in AO Images}
\begin{threeparttable}
\label{tab:ao}
\begin{tabularx}{\textwidth}{XXXXXXX}
\hline
EPIC      & $\rho$ & P.A.  & $\Delta m$ & Band & Epoch$^{\ddagger}$  & Ref.$^{\dagger}$ \\
          & (mas)  & (deg) & (mag)                  &      & (MJD)  &  \\
\hline
203843911 & $3697.3\pm1.8$ & $357.308\pm0.027$   & $8.737\pm0.112$  & $K_{\rm p}$+C600 & 56116.35 & TW \\
203770559 & $1491\pm20$      & $262.9\pm0.1$     & $0.838\pm0.040$  & $K_{\rm s}$      & 52094    & R05 \\
203895983 & $296.23\pm1.53$  & $70.203\pm0.290$  & $0.109\pm0.006$  & $K_{\rm p}$      & 57195.38 & TW  \\
203936815 & $1438\pm12$      & $69.5\pm0.3$      & $1.311\pm0.022$  & $K_{\rm s}$      & 51713    & R05 \\
203950167 & $1900\pm100$     & $38.4\pm1.0$      & 2.70             & $K$              & ...      & M10 \\
204211116 & $3947.6\pm2.7$   & $205.955\pm0.037$ & $7.240\pm0.032$  & $K_{\rm p}$      & 57584.38 & TW \\
204211116 & $3883.3\pm6.9$   & $350.469\pm0.101$ & $8.364\pm0.131$  & $K_{\rm p}$      & 57584.38 & TW \\
204489514 & $5378.5\pm0.6$   & $49.60\pm0.01$    & $4.61\pm0.02$    & $K_{\rm p}$      & 57169    & B19 \\
204489514 & $3636.2\pm4.8$   & $86.759\pm0.074$  & $8.064\pm0.086$  & $K_{\rm p}$      & 57584.33 & TW \\
205238942 & $4176.0\pm2.2$   & $171.346\pm0.028$ & $4.332\pm0.004$  & $K_{\rm p}$      & 57225.29 & TW \\
\hline
\end{tabularx}
\begin{tablenotes}
\item[$^{\ddagger}$] The starting epochs of the {\it K2}/C2 and {\it K2}/C15 light curves shown in Figure~\ref{fig:data} are MJD 56892 and 57988, respectively.
\item[$^{\dagger}$] B19$=$\cite{Barenfeld2019}; M10$=$\cite{McClure2010}; R05$=$\cite{Ratzka2005}; TW = This Work.
\end{tablenotes}
\end{threeparttable}
\end{table*}

\subsection{Disc Morphologies}
\label{sec:dsharp}

Figure~\ref{fig:data} shows that dipper discs exhibit a wide range of morphologies, from compact discs (e.g., EPIC~203937317) to extended discs (e.g., EPIC~203843911) to discs with large inner cavities (e.g., EPIC~204638512), and even discs with azimuthal asymmetries (e.g., EPIC~204281213). Although many dipper discs appear featureless at the current spatial resolution, the growing number of discs with very high spatial resolution ALMA data \cite[i.e., at scales of a few au;][]{HLTau2015, Andrews2016, Andrews2018} suggest that most discs will exhibit structure \cite[most commonly concentric rings and gaps;][]{Huang2018} if observed at sufficiently high spatial resolution.

Four of the dippers in our sample overlap with DSHARP, a programme that mapped the millimeter continuum of 20 protoplanetary discs at spatial resolutions of $\sim$5~au \citep{Andrews2018}. The DSHARP images of these four dippers are shown in Figure~\ref{fig:dsharp}. \cite{Huang2018} performed a systematic analysis of the annular substructures (i.e., the bright and dark annuli) visible in the DSHARP discs, finding a range of morphologies and no clear trends in disc architecture within their sample. With regards to the four dippers in our sample, they found that EPIC~204281213 (HD~143006) hosts a disc with complex structure, exhibiting three bright rings (centred at 6, 41, and 65~au) and two gaps (centred at 22 and 51~au) as well as a bright crescent at 80\degrees{} $< \theta <$ 144\degrees{} and potential inner cavity. In stark contrast, EPIC~203860592 (WSB~52) hosts a relatively compact disc with only one low-contrast ring and gap (centred at 25~au and 21~au, respectively) as well as an optically thick core that extends out to nearly 30~au.  The disc around EPIC~204514548 (HD~142666) shows four bright rings (centred at 6, 20, 40, and 58~au) and three gaps (centred at 16, 37, and 55~au) as well as a potential inner cavity surrounded by an inner disc brightness asymmetry, which may be due to viewing the heated and puffed-up interior of a ring. Finally, EPIC~203843911 (DoAr~25) hosts a particularly extended disc with three faint rings (centred at 86, 111, and 137~au) and three gaps (centred at 74, 98, and 125~au) around a bright (but not very optically thick) core. 

The morphological diversity of these four dipper discs appears to echo that seen in the overall DSHARP sample, with no obvious shared traits.  Just as many exhibit potential inner cavities as those that do not (although the sample size is small), and while all have some sort of ringed structure, this seems to be a common feature among the general disc population when observed at sufficiently high spatial resolution. Moreover, the morphology of the ringed structure is diverse among the dipper sample, reflecting what is also found in the larger DSHARP sample. We note that four other DSHARP discs (Elias~20, Elias~24, Elias~27, and AS~205) have {\it K2}/C2 light curves but do not exhibit dipper behavior.

\subsection{Stellar Multiplicity}
\label{sec:mult}

Many stars are in multiple systems \cite[see review in][]{DK2013} and blending in the {\it K2}/C2 light curves is a concern due to the large {\it Kepler} pixel sizes (4\as{} $\times$ 4\as{} or 520~au $\times$ 520~au at the distance of our sample). This risks complicating the interpretation of our inclination results, if more than one component could be the source of the dipper signal in the {\it K2} data or could host a disc that remains undetected and/or unresolved in our ALMA observations. Fortunately, all but two of the dippers in our sample have been surveyed for close companions with high-contrast imaging (see Section~\ref{sec:ao}). Eight were found to have candidate close ($\la$5\as{}) companions; Table~\ref{tab:ao} gives the separations ($\rho$), position angles (P.A.), and contrasts ($\Delta m$) of the detected candidate companions. For all sources in our sample with high-contrast imaging, detection limits as a function of separation from the primary, when available, are given in Table~\ref{tab:contrasts}. Moreover, we can use information from the {\it Gaia} Data Release 2 (DR2) as an independent check of binarity and also to assess the likeliness of any candidate companions being physically bound. The latter could be of interest when interpreting the dipper phenomenon and has implications as to whether the potential contaminants remain blended in the {\it K2} light curves taken at different epochs.

The imaged candidate companions to four of the dippers in our sample (EPIC~203843911, 204211116, 204489514, 205238942) are sufficiently separated and/or faint that they may be older background objects (e.g., see discussion in \citealt{Barenfeld2019}) and indeed we do not detect discs at the locations of their candidate companions in our ALMA data. The {\it Gaia} DR2 proper motions and parallaxes of EPIC~204489514 and its brighter candidate companion do not match and thus these sources are likely unassociated (the fainter candidate companion is not detected in {\it Gaia} DR2). Although EPIC~205238942 and its candidate companion have similar parallaxes and proper motions, making them possible true companions, the secondary's faintness rules it out as the source of the dipping (but also makes it an interesting potential low-mass brown dwarf companion of $\sim$30-50~M$_{\rm Jup}$ at $\sim$600~au). The candidate companions to EPIC~204211116 and 203843911 are not detected in {\it Gaia} DR2, likely due to their faintness, and their large separations make them unlikely to be associated to the dippers. 

Three of the remaining imaged candidate companions (to EPIC~203770559, 203936815, 203950167) are sufficiently close that they may be gravitationally bound, but also sufficiently separated that their discs may have avoided substantial tidal truncation \cite[e.g.,][]{Harris2012}, and indeed we detect discs around both components in two of these cases (EPIC~203770559 and 203936815). We do not detect a disc around the secondary to EPIC~203950167, and {\it Gaia} DR2 gives marginally inconsistent parallaxes (3--4$\sigma$ differences) and similar but inconsistent proper motions (1--3~mas~yr$^{-1}$ difference); regardless, the faint companion, although blended in the {\it K2}/C2 light curve, is unlikely to be the source of the dipping, as the photometric variability is larger than the primary-secondary flux ratio. 

The remaining dipper with an imaged candidate companion (EPIC~203895983) is in a close, nearly equal-mass system. Although we do not clearly resolve two discs in our ALMA data, the imaged disc is noticeably asymmetric compared to our overall sample (see Figure~\ref{fig:data}) and the elongation is along the same P.A. as the binary. This suggests that there are two discs that are blended in our ALMA data and that the derived inclination should not be trusted.

An independent check for binarity comes from the {\it Gaia} DR2 astrometric fits \citep{GaiaDr2} through the re-normalised unit weight error (RUWE). RUWE measures the goodness-of-fit (similar to reduced $\chi^2$) of the {\it Gaia} DR2 astrometric solution compared to stars of similar color and brightness. Large RUWE values ($>$1.4) are indicative of an unresolved companion impacting the solution to the DR2 astrometry \citep[e.g.,][]{Ziegler2019}. The two dippers in our sample lacking AO data (EPIC~204142243 and 205345560) have only moderate RUWE values (0.93 and 1.26, respectively) consistent with being single stars. EPIC~204638512 is the only dipper in our sample with ${\rm RUWE} > 1.4$, suggesting a companion too close and/or too faint to be detected by the current AO data (see Table~\ref{tab:contrasts}). Interestingly, this dipper hosts a disc with a large inner cavity whose properties are consistent with being carved by a massive planet \cite[e.g., ][]{Pinilla2018}. Although nebulosity around young stars could also increase their RUWE values \cite[e.g.,][]{Long2019}, EPIC~204638512 is located in the evolved Upper Sco region and has no evidence for surrounding nebulosity. Finally, EPIC~203895983 is the one star in our sample without a {\it Gaia} DR2 parallax, which is also a potential indicator of binarity, as parallaxes are only reported if a single-star solution is found; indeed, as discussed above, this source is a near-equal-mass binary in the AO data with a likely blended binary disc in the ALMA data.

Thus the dippers in our sample that may be of concern from a multiplicity standpoint are EPIC~203770559 and 203936815, for which we detect both discs in our ALMA observations, and EPIC~203895983, whose disc components are likely blended in our ALMA data. We flag these systems in the remaining analysis throughout this paper.


\section{Discussion}
\label{sec:discussion}

\subsection{Outer Disc Inclination Distribution}
\label{sec:incdist}

\begin{figure}
\begin{center}
\includegraphics[width=8cm]{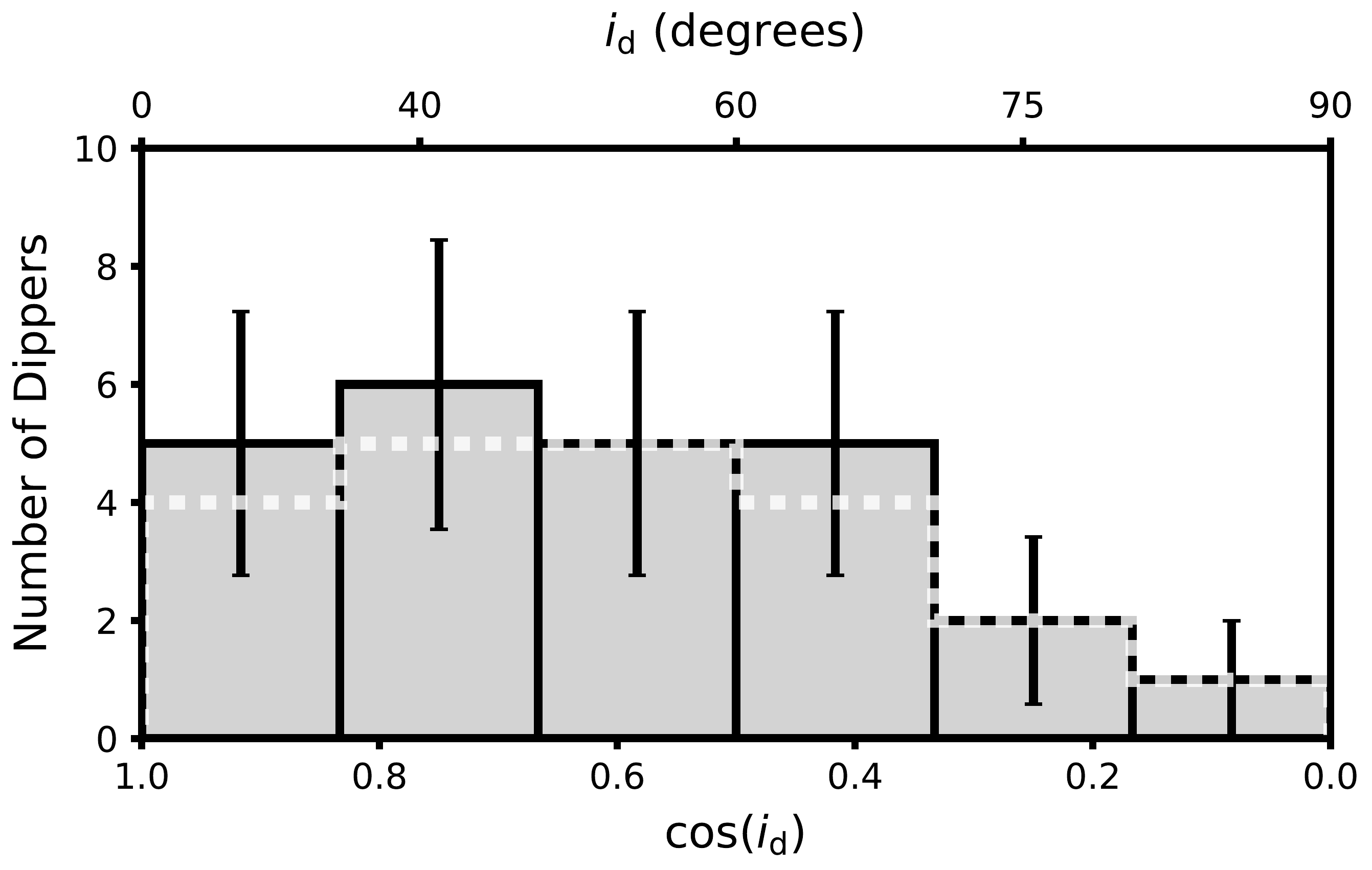}
\caption{The inclination distribution of dipper discs resolved by ALMA, where an isotropic distribution is flat in cos($i_{\rm d}$). The disc inclination values are taken from Table~\ref{tab:results}. For the two resolved binary disc systems, we use the inclination of the disc around the primary. The error bars are the square root of the number of dippers in the histogram bins. The white dashed line shows how the results would change if the two resolved binary discs and likely blended binary disc are removed (see Section~\ref{sec:mult}).}
\label{fig:inc_dist}
\end{center}
\end{figure}

Figure~\ref{fig:inc_dist} shows the distribution of dipper disc inclinations resolved by ALMA (Section~\ref{sec:discinc}; Table~\ref{tab:results}). For the two resolved binary discs (EPIC~203770559 and 203936815), we use the inclination of the disc around the primary, but also show the inclination distribution when removing these sources as well as the likely blended binary disc (EPIC~203895983) from the sample. In either case, the distribution appears approximately uniform with $\cos i_{\rm d}$ out to $i_{\rm d}\approx75$\degrees{}, suggesting an isotropic disc inclination distribution, with the exception of the most highly inclined cases.

The deficit of highly inclined systems is an expected observational bias, as these edge-on discs will obscure their host stars with their optically thick midplanes, making the stars too faint to be included in the {\it K2} target catalog. To confirm this quantitatively, we generated a grid of protoplanetary disc models using MCFOST \citep{Pinte2006}, a 3D Monte Carlo radiative transfer code that simulates images of discs at a given wavelength for specified disc structure and dust grain properties (see Appendix~\ref{appendix-B} for details of the model grid). Using our MCFOST model grid, we confirm that $\gtrsim50$~percent of stars with discs of $i_{\rm d}\gtrsim80$\degrees{} would be undetected by {\it K2}, rising to $\gtrsim90$~percent at $i_{\rm d}\gtrsim85$\degrees{} (Angelo et al., in prep), consistent with Figure~\ref{fig:inc_dist}. This supports the interpretation that the low dipper occurrence rate at high disc inclination is an observational bias.

\begin{figure}
\begin{center}
\includegraphics[width=8.3cm]{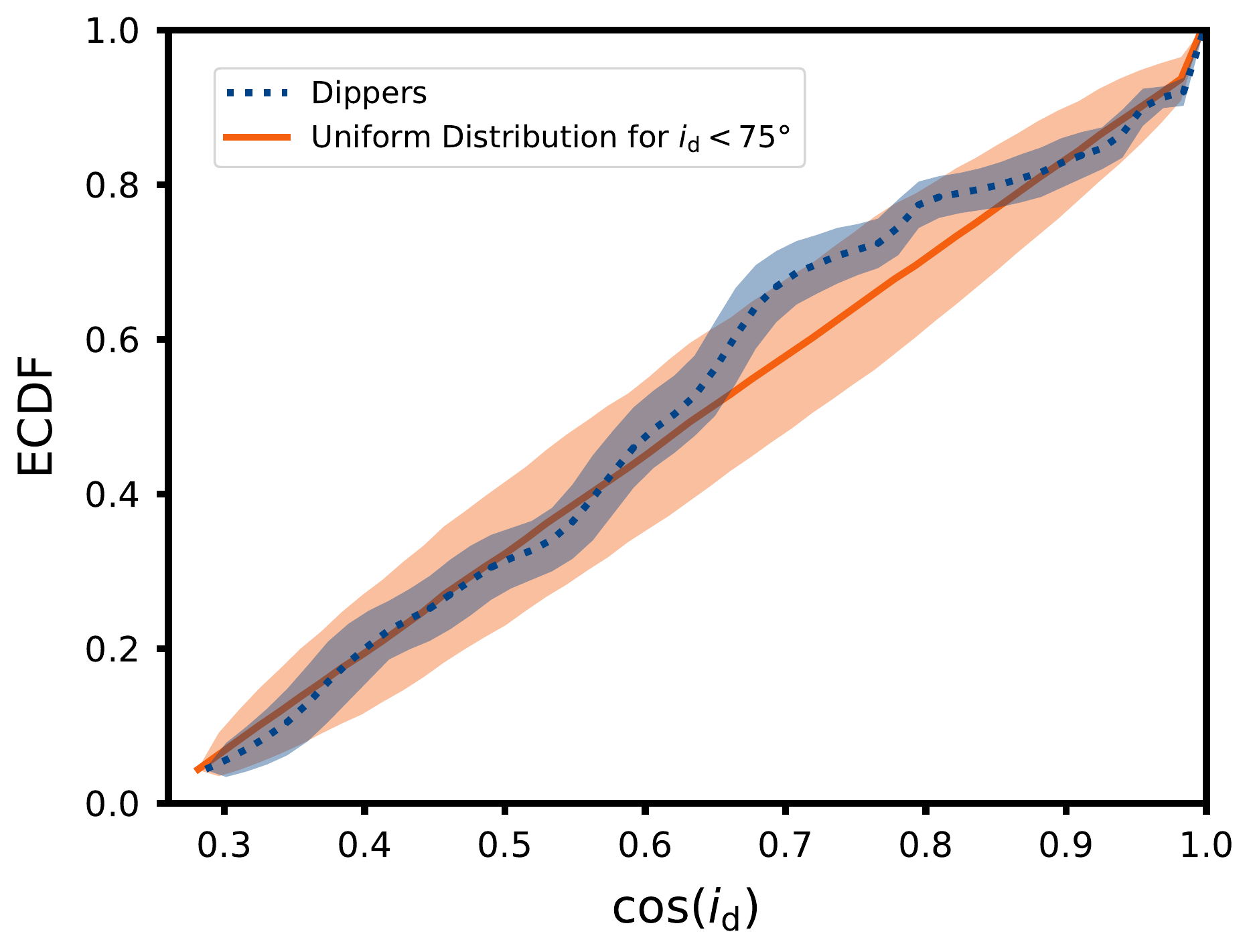}
\caption{The empirical cumulative distribution function (ECDF) of the dipper disc inclinations in our sample (blue dashed line) compared to a uniform distribution for $i_{\rm d}<75$\degrees{} (solid orange line); the shaded regions represent 1$\sigma$ uncertainties (see Section~\ref{sec:incdist}). We only consider $i_{\rm d}<75$\degrees{} for the uniform distribution as more edge-on discs will block their star, creating an observational bias seen in the observed dipper sample (see Section~\ref{sec:incdist} and Figure~\ref{fig:inc_dist}). These two distributions being statistically indistinguishable suggests that dippers have an isotropic disc inclination distribution.}
\label{fig:inc_ecdf}
\end{center}
\end{figure}

In order to more robustly test whether the inclination distribution at $i_{\rm d}\lesssim75$\degrees{} is consistent with isotropic, we construct an empirical cumulative distribution function (ECDF), which is a non-parametric estimator of the cumulative distribution function for a random variable. We build the ECDF by randomly sampling from the inclination posterior distributions inferred with {\tt GALARIO} for each dipper, converting these to cos($i_{\rm d}$), then applying the {\tt ECDF} package in the {\tt STATSMODEL} Python module. We then repeat this 1000 times and take the mean and standard deviation as the final values and associated uncertainties of the ECDF. The result is shown in Figure~\ref{fig:inc_ecdf} and compared to the isotropic case, constructed by randomly sampling from a uniform distribution between 0.26 and 1.0 for cos($i_{\rm d}$) (corresponding to $0 < i_{\rm d} < 75$\degrees{} due to the observational bias discussed above) for each dipper in our sample (i.e., the sample sizes are the same). The distributions clearly overlap, demonstrating that the dipper disc inclination distribution is consistent with isotropic. The average and standard deviation of the p-values of two-sided Kolmogorov-Smirnov tests, calculated for each of the 1000 draws and only considering cos($i_{\rm d}$)$>0.26$, is 0.64$\pm$0.27, indicating that we cannot reject the null hypothesis that the dipper cos($i_{\rm d}$) distribution and the uniform distribution are drawn from the same parent distribution (this result holds even when removing the three binary discs).

Our findings are in contrast to those of \cite{Cody2018}, who reported that dipper discs favor higher inclinations ($i_{\rm d}\gtrsim50$\degrees{}) apart from a few face-on exceptions. They used a similarly selected sample (i.e., dippers identified by their {\it K2}/C2 light curves with disc inclinations derived from resolved ALMA data), however their disc inclinations were taken directly from the literature rather than being uniformly analyzed and no formal statistical tests were reported. Moreover, many of the highly inclined discs used in \cite{Cody2018} were taken from \cite{Barenfeld2017} and have particularly large uncertainties ($\pm $40--60\degrees{}). In Appendix~\ref{appendix-C}, we present these dipper discs with large uncertainties and show that including them in our samples only makes the dipper disc inclination distribution even more consistent with isotropic by filling in the high inclination end of the distribution. 

Finally, we note that four sources in our sample are A-type or G-type stars (EPIC 204281213, 203950167, 204399980 and 204514548; see Table~\ref{tab:results}). Although earlier-type stars exhibiting dimming events are often classified as UXOR variables, the photometric variability of dippers and UXORs are distinctly different: UXORs are characterized by deep (up to several magnitudes), long-term (weeks to years) dimming events while dippers are characterized by shorter (days) and relatively shallower (a few tens of percent) dimming events. Moreover, there are now several examples of young stars exhibiting both dipper and UXOR phenomenona over time \cite[namely the prototypical dipper AA~Tau, which is currently undergoing a UXOR-type dimming event;][]{Bouvier2013}. Nevertheless, removing these four sources from our sample does not change our overall results, as their disc inclinations are evenly distributed.

\subsection{Correlations with Light Curve and Disc Properties}
\label{sec:discprop}

\begin{figure}
\begin{center}
\includegraphics[width=8cm]{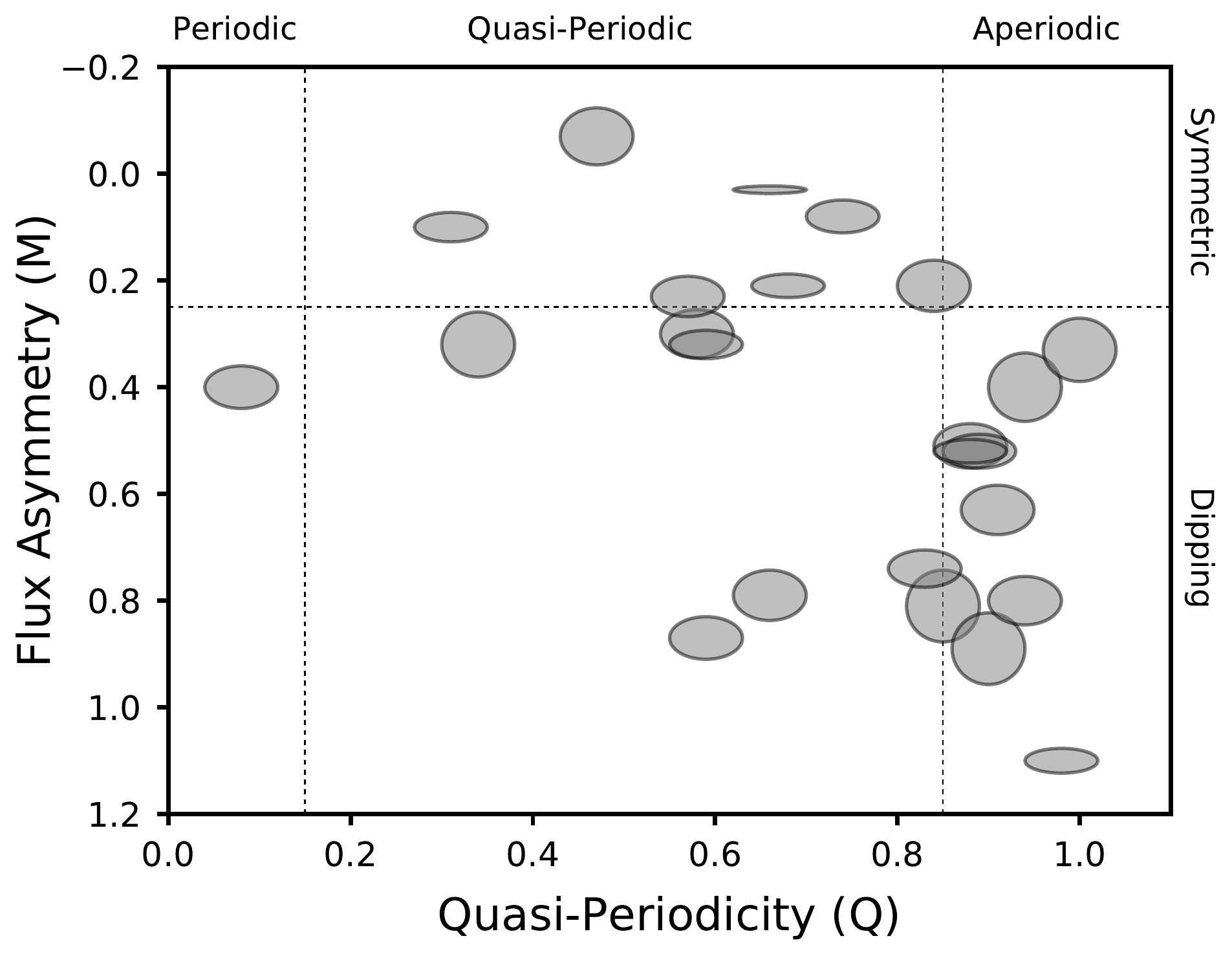}
\caption{Graphical representation of dipper disc inclinations, distributed in a plot of flux asymmetry ($M$) versus quasi-periodicity ($Q$), as defined in \citet{Cody2014} and calculated in \citet{Cody2018}. The dashed lines are the delimiters of the different types of light curve morphologies, identified by eye in \citet{Cody2018}. There is no perceptible pattern in dipper disc inclination with variability type (see Section~\ref{sec:incdist}).}
\label{fig:qm}
\end{center}
\end{figure}

Different mechanisms to explain the dippers (e.g., accretion-related inner disc warps vs. disc winds; see Section~\ref{sec:theory}) may produce light curves with different morphologies and be more likely to occur at different disc inclinations. Figure \ref{fig:qm} therefore graphically represents the dipper disc inclinations, plotted in the light curve morphology space, as represented by the flux asymmetry ($M$) and quasi-periodicity ($Q$) statistics defined in \citet{Cody2014}. According to these statistics, light curves with symmetric flux distributions about a median amplitude (e.g., sinusoidal star spot patterns) have $M=0$, and more negative-going light curves (e.g., dippers) have higher positive $M$ values; perfectly periodic light curves have $Q=0$, while those with $Q\approx1$ are aperiodic. Figure~\ref{fig:qm} illustrates that there is no perceptible correlation between disc inclination and light curve morphology, at least as defined by the $Q$ and $M$ statistics of \citet{Cody2014}. The lack of low $Q$ and high $M$ sources (i.e., quasi-periodic dippers with large negative deviations relative to the median) should not be interpreted as the quasi-periodic dippers tending to have shallow dips; rather, this is likely due to quasi-periodic dippers often being dominated by star-spot patterns with high duty cycles (e.g., EPIC~ 203860592, furthest left in Figure~\ref{fig:qm}), which drives the median of the light curve to more negative values, and thus the $M$ statistic to smaller positive values. 

One explanation for this lack of correlation is that the photometric behavior of dippers is known to change over year-long (and possibly shorter) timescales. \citet{McGinnis2015} found that some dippers in NGC~2264, observed by {\it CoRoT} in 2008 and then again in 2011, switched between aperiodic and quasi-periodic variability (or vice versa) at some point between the two epochs (and possibly more than once). Significant changes in dipper light curve morphology on similar timescales are also seen when comparing the {\it K2}/C2 and {\it K2}/C15 light curves in Figure~\ref{fig:data} for the three sources in our sample that were observed in both campaigns, which were separated by three years. In contrast, the prototypical dipper AA Tau maintained a clear quasi-periodic dipping pattern for at least 20 years \citep{Bouvier2013}. Building up a larger sample of dippers with multiple epochs of high-precision photometry are needed to investigate this possibility, and this should become possible in the near future, for example by combining the {\it K2} and Transiting Exoplanet Survey Satellite \citep[{\it TESS};][]{Ricker2014} datasets.

\begin{figure}
\begin{center}
\includegraphics[width=8cm]{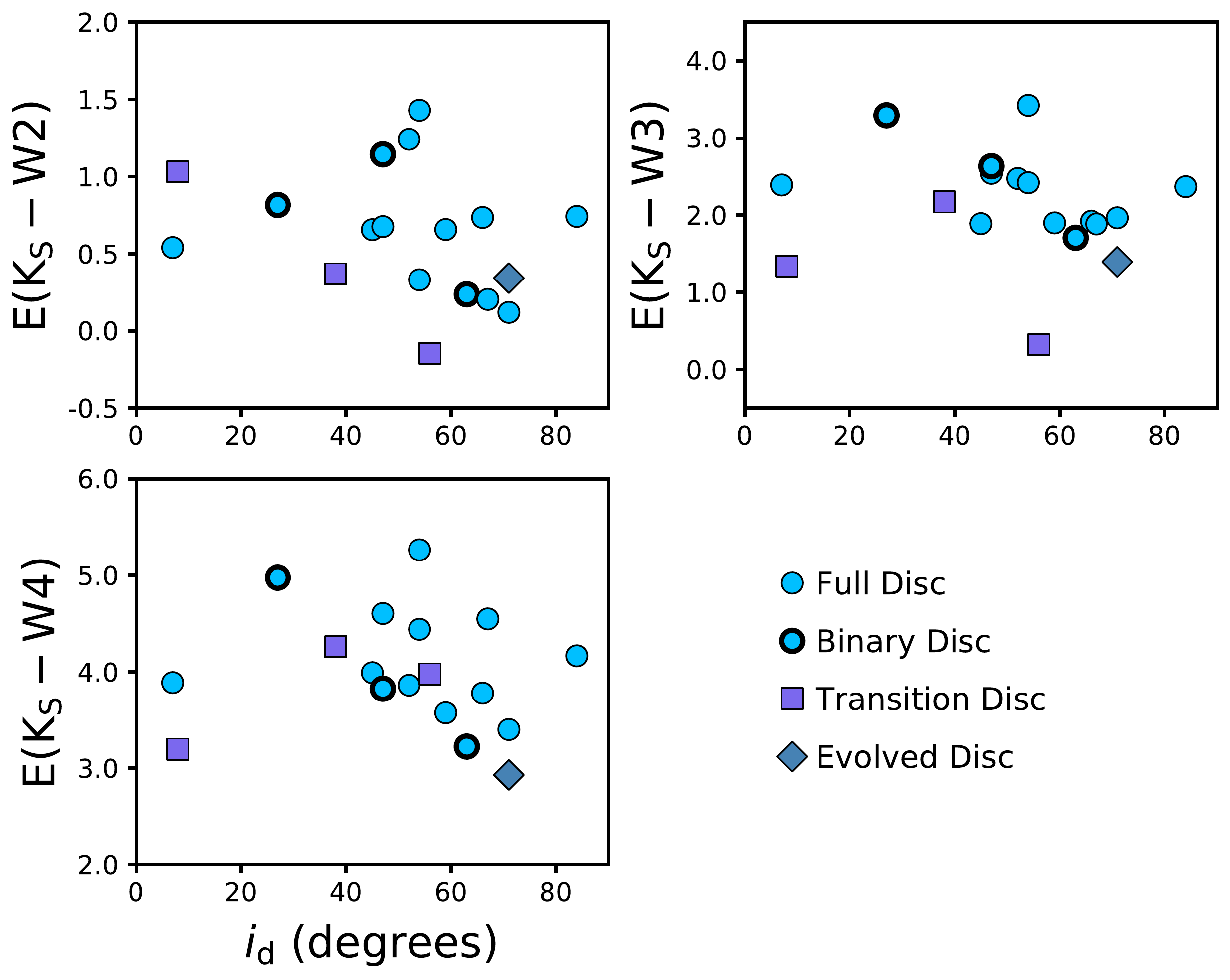}
\caption{{\it WISE} mid-infrared excess over the stellar photosphere versus disc inclination for the late-type (K/M) dippers in our sample (see Section~\ref{sec:discprop}). Circles are ``full" discs, squares are ``transition" discs with large inner dust cavities, and diamonds are ``evolved" discs. Those with thicker outlines are the three binary discs discussed in Section~\ref{sec:mult}.}
\label{fig:wise}
\end{center}
\end{figure}

To investigate correlations between disc properties and disc inclination, we plot mid-infrared excesses (relative to the expected stellar photosphere values) in the Wide-field Infrared Survey Explorer \cite[{\it WISE};][]{Wright2010} W2~(4.6$\mu$m), W3~(12$\mu$m), and W4~(25$\mu$m) bands against $i_{\rm d}$ in Figure~\ref{fig:wise}. These excesses are calculated as in \cite{LM2012}, and we also follow their procedure for using these excesses to classify disc type \cite[e.g., full, transitional, evolved; see Figure~6 in][]{Ansdell2016a}. We note that these methods use spectral type as a reference for the stellar photosphere, and that the spectral types used in this work are taken from inhomogeneous literature sources (see Table~\ref{tab:results}), thus this analysis should be repeated once homogenous spectral types are derived for our sample. The W2 wavelength corresponds to peak blackbody emission at $\sim$600~K, which is the expected temperature of dust grains orbiting at $\sim$10 stellar radii around these pre-main sequence stars, while the W3 and W4 emission probe cooler dust within an au to a few au. Thus the amount of {\it WISE} excess might be expected to depend on disc viewing geometry: all else being equal, a highly inclined, optically thick disc will subtend a smaller solid angle and produce a smaller mid-infrared excess. Moreover, at very high inclinations, the inner regions of a flared disc will become (partially) obscured, attenuating emission at near-infrared wavelengths. 

\begin{figure*}
\begin{center}
\includegraphics[width=16.8cm]{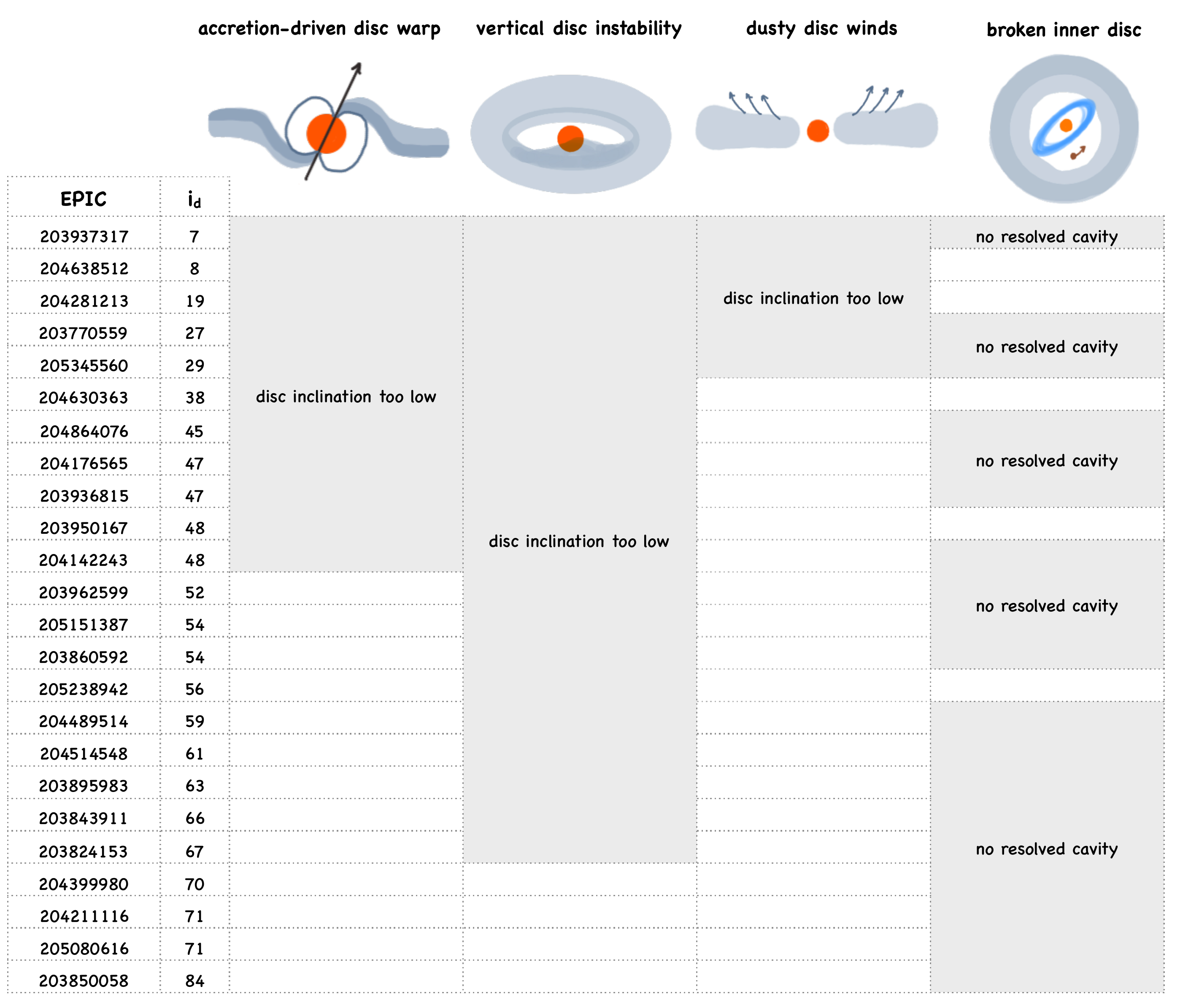}
\caption{Proposed dipper mechanisms and their feasibility for the dippers in our sample given our derived outer disc inclinations. Note that for the ``accretion driven inner disc warp" sceneario, we consider inclined stellar magnetic fields that allow for more moderate disc inclinations \citep{Bodman2017}. Moreover, for the ``broken inner disc" scenario, future ALMA observations at higher spatial resolution may reveal smaller inner cavities not yet resolved.}
\label{fig:mech}
\end{center}
\end{figure*}

We find no statistically significant correlations between disc inclination and {\it WISE} excess (Spearman rank tests return $\rho=-0.37$, $-0.31$, and $-0.25$ with $p-$values of 0.13, 0.21, and 0.32 for the W2, W3, and W4 excesses, respectively). Nonetheless, Figure \ref{fig:wise} does hint toward decreasing W4 excess with higher disc inclinations, as expected from the above geometric arguments. This suggests that disc material at separations ($\sim$2~au) and temperatures ($\sim$100K) that correspond to peak emission at 25~$\mu$m is co-inclined with the outer ($\gtrsim$10~au) disc resolved by ALMA. The two exceptions have the lowest inclinations: EPIC~20397137 (DoAr~24), a very compact disc, and EPIC~204638512 (J1604), a transition disc with a thin annulus, both of which may lack material at $\sim$2~au. The absence of a correlation between W2 excess and disc inclination may be explained by the structure of the inner ($\lesssim$1~au) disc being completely unrelated to the geometry of the outer disc resolved by ALMA. Moreover, emission at these shorter wavelengths can be variable: a striking example is EPIC~204638512 (J1604), whose {\it Spitzer}/IRAC photometry (taken in the early 2000s) shows no infrared excess at $\lesssim$10~$\mu$m, while its {\it WISE} photometry (from 2010) and {\it Spitzer}/IRS spectra (from 2007) show significant excess at these same wavelengths \cite[e.g., see Figure 4 in][]{zhang2014}.

\subsection{Expectations from Theory}
\label{sec:theory}

Multiple mechanisms have been considered to explain the dippers, and they often require or are biased toward specific viewing geometries. Therefore in this section, we compare the expectations from these theories to our observed disc inclination distribution, as summarized in Figure~\ref{fig:mech}.

One of the first mechanisms that was crafted to explain the dipper phenomenon---and in particular the 8.5-day quasi-periodic photometric variability of the prototypical dipper AA Tau---is highly non-axisymmetric, magnetically funneled accretion from the inner disc edge onto the star, creating an inner disc warp that partially occults the star as the disc rotates \citep{Bouvier1999}. This ``accretion-driven inner disc warp" scenario requires nearly edge-on disc inclinations \citep[$\sim$70\degrees{}; e.g., see][]{McGinnis2015, Kesseli2016}, a viewing geometry supported for AA~Tau by studies of polarization \citep{Menard2003} and emission line shape \citep{Bouvier2003}, though disputed by ALMA observations that clearly show a moderately inclined outer disc \cite[59.1$\pm$0.1\degrees{};][]{Loomis2017}. This mechanism could still explain the dipping behavior of AA~Tau, however, as \cite{Bodman2017} used magnetospheric truncation theory to show that, if the stellar magnetic field axis is sufficiently tilted, this mechanism could be extended to moderate disc inclinations. However, unlike AA~Tau, most dippers have accretion rates that are below the classical T~Tauri level \citep{Ansdell2016a}, making the accretion-driven disc warp scenario unlikely in most cases, especially for the aperiodic dippers that require high accretion rates to drive unstable accretion regimes \citep{KR2013}. Moreover, it is telling that of the four most highly inclined ($i_{\rm d}\gtrsim70$\degrees{}) dippers in our sample, only EPIC~204211116 has an AA Tau-like (i.e., quasi-periodic) light curve, while there are multiple examples at moderate inclinations (see Figure \ref{fig:data}). 

Another explanation for the dippers that requires nearly edge-on viewing geometries is occultations of the star by vertical structures produced by some instability in the disc. One incarnation of this ``vertical disc instability" scenario involves vortices produced by Rossby waves, which occur at an extremum in the disc vortensity \citep[i.e., the ratio of the vorticity to density;][]{Lovelace1999,Meheut2012,Meheut2012b,Meheut2013,Lin2013,Ono2016}, such as at the boundary of a ``dead zone'' where the magnetorotational instability ceases to operate \citep{Lyra2012,Miranda2016,Miranda2017}. These structures will be limited to 1--2 scale heights above the disc; at the characteristic temperatures ($\sim$1000~K) and orbital periods ($\sim$4~days) where the structures would have to be located to explain the dippers \cite[i.e., assuming $\sim$10~Myr old K/M type stars with the occulting dust orbiting near the star-disc co-rotation radius;][]{Ansdell2016a,Bodman2017}, the scale height would be $\sim$0.1~au and thus the disc inclinations would be limited to $i_{\rm d}\gtrsim70$--80\degrees{}. This rules out the vertical disc instability scenario for most of the dippers, if the inner disc where these instabilities occur has the same geometry as the outer disc resolved by ALMA.

One mechanism for the dippers could involve dust being lofted in disc winds driven by stellar XUV radiation and/or impelled by radial magnetic fields \citep{BP1982}. Dust clouds crossing our line of sight can contribute to both photometric variability and a ``bump" in infrared emission near 3~\micron\ \citep{Tambovtseva2008,Bans2012}. This ``dusty disc wind" scenario could also explain emission from silicates well above the disc mid-plane \citep{Varga2017,Giacalone2019}, the infrared variability of MWC~480 \citep{Fernandes2018}, and contemporaneous optical dimming and infrared brightening of HD~163296 \citep{Ellerbroek2014}. However, magnetised disc winds are limited to angles $\gtrsim30$\degrees{} from pole-on \citep{BP1982}, and if we assume the occultation occurs near the wind launch radius, then we should expect a deficit of nearly face-on dipper discs. This is also not seen in our observed distribution, although obscuration by associated jets could be responsible for systems with the lowest inclinations.

One possibility, of course, is that there are significant misalignments between the outer discs resolved by ALMA and the unseen inner discs. Directly constraining the geometry of inner discs is difficult as it requires near-infrared interferometry and thus is only possible for bright (and thus typically high-mass) sources. There are a couple of bright UXOR objects with resolved outer disc geometries whose inner ($<$1~au) disc inclinations have also been constrained by near-infrared interferometry: the archetype UX~Ori has an inner and outer disc inclination of $\sim$70\degrees{} revealed by near- and mid-infrared interferometry, respectively \citep{Kreplin2016}; CQ~Tau has an inner disc inclination of $\sim$48\degrees{} constrained by near-infrared interferometry \citep{Eisner2004} while ALMA measures an outer disc inclination of $\sim$37\degrees{} \citep{Pinilla2018a}. For these sources, the inner and outer discs appear fairly aligned. However, this is expected for UXORs, as their variability has been successfully explained by puffed-up inner disc rims that also self-shadow their outer discs, causing their observed weak far-infrared excesses \citep{Dullemond2003}. Unfortunately, most inner discs cannot be resolved for the typically late-type (and thus faint) dipper stars, although one of the higher-mass dippers in our sample, EPIC~204514548 (HD~142666), has an outer disc inclination of 61\degrees{} measured by ALMA and an inner disc inclination of $\sim$58\degrees{} constrained with CHARA \citep{Davies2018b}. Misalignments can be indirectly detected, however, via the shadows that an inclined inner disc casts on the outer dust disc in high-contrast optical/infrared images \cite[e.g.,][]{Marino2015,Min2017,Pinilla2018} or by velocity perturbations imprinted in the gas kinematics \cite[e.g.,][]{Teague2018}. Velocity perturbations in HCO$^+$ emission have been seen for the dipper AA~Tau \citep{Loomis2017}, and both kinematic signatures in the gas \citep{Mayama2018} and variable shadows in scattered light \citep{Takami2014, Pinilla2018} have been observed for EPIC~204638512 (J1604). For the latter, \cite{Davies2019} also derive a stellar inclination that is highly misaligned with the outer disc.

These misaligned inner discs could be induced by stellar companions \cite[e.g.,][]{Facchini2018} or planets with masses down to $\sim1~M_{\rm Jup}$ \cite[e.g.,][]{MK2017,Zhu2019,Nealon2019} on orbits inclined relative to the outer disc. Simulations show that when these companions open a gap, the disc inside the orbit breaks from the outer regions and becomes misaligned. The inner discs can be highly misaligned with respect to the outer disc \citep[i.e., $\gtrsim$70\degrees{};][]{Min2017,Facchini2018}, producing pairs of narrow shadows in the outer disc that can be highly dynamic. Indeed, \cite{Pinilla2018} observed with VLT/SPHERE \citep{Beuzit2019} that a pair of narrow shadows on the outer disc around one of the face-on dippers in our sample, EPIC~204638512 (J1604), were variable both in morphology and in position on timescales of days. Even moderately inclined planet orbits ($\sim$10\degrees{}) can misalign inner discs and cast shadows in the outer regions \cite[e.g.,][]{Nealon2019}, although these shadows should be broad in extent (rather than narrow lanes), and could explain discs observed to be covered up to half in shadow \cite[e.g.,][]{Benisty2018}. Thus this ``broken inner disc" scenario could explain dippers with a range of outer disc inclinations, including those with low or moderate outer disc inclinations, although it requires a stellar or massive planetary companion orbiting in the inner disc. Our AO imaging can rule out stellar (but not planetary) companions at a few au around some dippers in our sample (see Table \ref{tab:contrasts}). Although stellar companions in the inner ($<$1~au) disc would not be resolved in our AO imaging, they could be detected as spectroscopic binaries. Moreover, even though many of the ALMA images presented here do not show inner gaps \cite[the rate of these ``transition" discs in our sample is similar to that seen in the general disc population; see][]{Ansdell2016c}, the spatial resolutions---even of the DSHARP sample ($\sim$5 au)---are insufficient to resolve such small inner cavities.

It is of course possible that more than one (or all) of these mechanisms are responsible for the dipper phenomenon. Figure~\ref{fig:mech} illustrates which mechanisms can be ruled out for the dippers in our sample based on the above discussion of disc geometry and currently available data.


\section{Conclusions} 
\label{sec:summary}

Dippers are a common class of young variable star often assumed to host protoplanetary discs viewed nearly-edge on, such that dusty structures lifted slightly out of the midplane partially occult the star as the disc rotates, producing the characteristic dimming events seen in their optical light curves. Until recently, it was difficult to robustly test this assumption of disc geometry due to the limited number of dippers with resolved discs. This has changed with the advent of ALMA and recent flurry of protoplanetary disc observations conducted at moderate ($\gtrsim$10~au) angular resolution \citep[e.g.,][]{Barenfeld2016,Cieza2018}.

Motivated by the earlier discovery of a dipper disc with a face-on geometry \cite[J1604;][]{Ansdell2016b}, we investigated the distribution of dipper disc inclinations resolved by ALMA. We found a disc inclination distribution consistent with isotropic over $i_{\rm d}\approx0-75$\degrees{} (with a deficit at higher inclinations being consistent with an observational bias due to optically thick disc midplanes blocking their host stars). We also found diverse disc morphologies on the $\gtrsim$10~au scales typically probed by our ALMA observations, also evident at $\sim$5~au scales for the four dippers observed at high spatial resolution as part of DSHARP \citep{Andrews2018}. 

These findings indicate that the dipper phenomenon is unrelated to the outer ($>$10~au) disc geometry probed by ALMA, and that any connection with disc morphology remains unclear. Given several lines of evidence that the dipper events are caused by dust in the inner ($<$1~au) disc, these findings further hint that inner disc misalignments may be common in protoplanetary discs around later-type stars. This interpretation is supported by recent results from high-contrast optical/infrared imagers that have revealed outer disc shadows likely cast by unseen misaligned inner disc components \cite[e.g.,][]{Debes2017,Benisty2018,Casassus2018,Pinilla2018}. Such misaligned discs would be distinct from UXOR systems, which are expected to have aligned inner and outer disc components in order to explain both their dimming events (caused by a puffed-up inner disc rim) and their weak far-infrared emission (due to self-shadowing of the outer disc by the inner disc rim) \citep{Dullemond2003}. Potential mechanisms causing the inner disc features hypothesized for the dippers include accretion-driven inner disc warping and ``breaking" of the inner and outer disc due to (sub-)stellar or planetary companions on inclined orbits.

There are several important avenues for future work. Higher contrast optical/infrared observations from instruments like SPHERE and GPI are needed to search for fainter companions, in particular among the transition discs whose inner cavities may be carved out by planetary-mass objects. Moreover, multi-epoch observations from such instruments will give insight into the occurrence and variability of shadows in the outer disc, and thus the presence of unseen misaligned inner disc components. High-resolution spectra, probably obtained in the infrared where there is more signal from these late-type and reddened stars, could identify spectroscopic binaries that can ``break" circumbinary discs. Inclinations of the rotational axis of the central stars (derived from measurements of $v \sin i$, stellar rotation period, and stellar radius) can also be compared to the outer disc inclinations, as any difference would further indicate misalignments in discs \citep[an initial attempt using literature values has been conducted by][]{Davies2019}. Finally, given that the photometric behavior and mid-infrared excesses of dippers are known to change on timescales of months to years, simultaneous multi-wavelength observations and long-term monitoring will be critical for better understanding this common class of young variable and thus the dynamic inner regions of protoplanetary discs more generally.


\section*{Acknowledgements}

MA gratefully acknowledges the NVIDIA Corporation for donating the Quadro P6000 GPU used for this research. 
MA acknowledges support from NSF grant AST-1518332 and NASA grants NNX15AC89G and NNH18ZDA001N/EW.
MA, GD, and IA acknowledge support from NASA grant NNX15AD95G/NEXSS.
MT is supported by the UK Science and Technology research Council (STFC).
JC acknowledges support from the National Aeronautics and Space Administration under grant No. 15XRP15 20140 issued through the Exoplanets Research programme. 
GMK is supported by the Royal Society as a Royal Society University Research Fellow. 
We thank Scott Barenfeld and Lucas Cieza for providing the calibrated ALMA visibilities of their disc survey programmes.
Part of this research was carried out at the Jet Propulsion Laboratory, Caltech, under a contract with NASA.
This paper includes data from the {\it Kepler} mission; funding for {\it Kepler} is provided by the NASA Science Mission directorate. Some data used in this paper were obtained from the Mikulski Archive for Space Telescopes (MAST) at the Space Telescope Science Institute (STScI). STScI is operated by the Association of Universities for Research in Astronomy, Inc., under NASA contract NAS5-26555.
We used data from the following ALMA programmes: 2011.0.00526.S, 2012.1.00046.S, 2013.1.00395.S, 2013.1.00498.S, 2015.1.01600.S, 2016.1.00336.S, 2016.1.00545.S, 2017.1.01180.S, and 2018.1.00564.S. ALMA is a partnership of ESO (representing its member states), NSF (USA) and NINS (Japan), together with NRC (Canada), NSC and ASIAA (Taiwan), and KASI (Republic of Korea), in cooperation with the Republic of Chile. The Joint ALMA Observatory is operated by ESO, AUI/NRAO and NAOJ. The National Radio Astronomy Observatory is a facility of the National Science Foundation operated under cooperative agreement by Associated Universities, Inc. 
This work has made use of data from the European Space Agency (ESA) mission {\it Gaia} ({\url https://www.cosmos.esa.int/gaia}), processed by the {\it Gaia} Data Processing and Analysis Consortium (DPAC, {\url https://www.cosmos.esa.int/web/gaia/dpac/consortium}). Funding for the DPAC has been provided by national institutions, in particular the institutions participating in the Gaia Multilateral Agreement. 
Some of The data presented herein were obtained at the W. M. Keck Observatory, which is operated as a scientific partnership among the California Institute of Technology, the University of California and the National Aeronautics and Space Administration. The Observatory was made possible by the generous financial support of the W. M. Keck Foundation.
The authors wish to recognize and acknowledge the very significant cultural role and reverence that the summit of Maunakea has always had within the indigenous Hawaiian community.  We are most fortunate to have the opportunity to conduct observations from this mountain.
We used the following software: {\tt lightkurve} \citep{LightKurve2018}, {\tt Astropy} \citep{AstroPy2013,AstroPy2018}, {\tt Matplotlib} \citep{Hunter2007}, {\tt uvplot} \citep{uvplot_mtazzari}, {\tt GALARIO} \citep{Tazzari2018}.


\bibliographystyle{mnras}

\appendix

\section{Imaging Detection Limits}
\label{appendix-A}

\afterpage{
\begin{landscape}
\begin{table}
\caption{Detection Limits from AO Imaging}
\begin{threeparttable}
\label{tab:contrasts}
\begin{tabular}{clcccccccccccccccl}
\hline
EPIC      & Epoch    & Band$^{d}$   & \multicolumn{14}{c}{Contrast $\Delta m$ (mag) at $\rho = $ (mas)} & Ref.$^{a}$ \\
          &          &              & 10   & 20   & 40   & 80   & 150  & 200  & 250  & 300 & 400 & 500 & 700 & 1000 & 1500 & 2000 & \\
\hline
203770559 & 52094    & $K_{\rm s}$  & ...  & ...  & ...  & ...  & ...  & ...  & ...  & ... & ... & ... & ... & ... & ... & ... & R05$^{b}$ \\
203824153 & 55674    & $K_{\rm p}$+NRM  & 0.00 & 1.53 & 2.86 & 2.69 & 2.69 & 2.61 & 2.74 & ... & ... & ... & ... & ... & ... & ... & C15$^{c}$ \\
203843911 & 55675    & $K_{\rm p}$+NRM  & 1.89 & 3.63 & 4.58 & 4.32 & 4.32 & 4.37 & 4.37 & ... & ... & ... & ... & ... & ... & ... & C15$^{c}$ \\
203843911 & 56116.35 & $K_{\rm p}$+C600 & ...  & ...  & ...  & ... & 3.7 & 4.1 & 4.3 & 4.6 & 7.2 & 7.5 & 8.7 & 10.1 & 11.2 & 11.9 & TW  \\
203850058 & 57584.34 & $K_{\rm p}$  & ...  & ...  & ...  & ...  & 4.8  & 5.3  & 6.1  & 6.3 & 6.9 & 7.1 & 7.8 & 8.2 & 8.2 & 8.3 & TW  \\
203860592 & 56360    & $K_{\rm s}$+NRM  & 1.64 & 2.20 & 4.23 & 3.24 & 3.24 & 4.00 & 4.00 & ... & ... & ... & ... & ... & ... & ... & C15$^{c}$ \\
203895983 & 57195.38 & $K_{\rm p}$  & ...  & ...  & ...  & ...  & 4.6  & 5.0  & 5.5  & 6.1 & 6.5 & 6.9 & 7.5 & 8.0 & 8.1 & 8.2 & TW  \\ 
203936815 & 51713    & $K_{\rm s}$  & ...  & ...  & ...  & ...  & ...  & ...  & ...  & ... & ... & ... & ... & ... & ... & ... & R05$^{b}$ \\
203937317 & 57195.34 & $K_{\rm p}$+C600 & ...  & ...  & ...  & ...  & 4.0  & 4.9  & 5.2  & 5.8 & 7.2 & 7.7 & 8.7 & 10.1 & 11.2 & 11.8 & TW \\
203950167 & ...      & $K$          & ...  & ...  & ...  & ...  & ...  & ...  & ...  & ... & ... & ... & ... & ... & ... & ... & M10$^{b}$ \\
203962599 & 52090    & $K_{\rm s}$  & ...  & ...  & ...  & ...  & 3.5  & ...  & ...  & ... & ... & 4.2 & ... & ... & ... & ... & R05 \\ 
204142243 & ...      & ...          & ...  & ...  & ...  & ...  & ...  & ...  & ...  & ... & ... & ... & ... & ... & ... & ... & ... \\
204176565 & 54635    & $K_{\rm p}$+NRM  & 3.12 & 4.62 & 5.45 & 5.33	& 5.33 & 5.28 & 5.33 & ... & ... & ... & ... & ... & ... & ... & C15$^{c}$ \\
204211116 & 57584.38 & $K_{\rm p}$+NRM  & ...  & 2.40 & 3.56 & 3.26 & 5.0  & 5.7  & 5.6  & 6.0 & 6.7 & 7.1 & 7.8 & 8.7 & 8.7 & 8.7 & TW  \\
204281213 & 54251.5  & $H$          & ...  & 3.49 & 5.06 & 5.43 & 5.43 & 5.35 & ...  & ... & ... & ... & ... & ... & ... & ... & K08$^{c}$ \\
204399980 & 55311.58 & $L^{\prime}$ & ...  & 2.66 & 4.42 & 4.83 & 4.72 & 4.78  & ... & ... & ... & ... & ... & ... & ... & ... & TW \\
204489514 & 57169.50 & $K_{\rm p}$+NRM  & 0.00 & 0.58 & 2.10 & 2.38 & 2.38 & 3.56 & 5.75 & 5.75 & 5.58 & 7.51 & 7.51 & 8.03 & 8.03 & 8.03 & B19$^{c}$ \\
204489514 & 57584.33 & $K_{\rm p}$+NRM  & ...  & 1.25  & 3.23  & 2.92 & 5.0 & 5.3 & 6.1 & 6.5 & 6.7 & 7.4 & 7.9 & 8.5 & 8.5 & 8.5 & TW \\
204514548 & 51621    & STIS         & ...  & ...  & ...  & ...  & ...  & ...  & ...  & ... & ... & ... & ... & ... & ... & ... & G05$^{b}$ \\
204630363 & 57225.30 & $K_{\rm p}$+NRM+C600  & 0.65  & 3.73  & 4.66  & 4.38  & 5.1  & 6.1  & 6.4  & 6.9 & 7.5 & 8.1 & 9.3 & 10.6 & 11.3 & 11.7 & TW  \\ 
204638512 & 54256.50 & $K_{\rm p}$+NRM  & 3.57 & 5.43 & 6.23 & 6.15 & 6.15 & 5.79 & 5.50 & 5.50 & ... & ... & ... & ... & ... & ... & K08$^{c}$ \\
204638512 & 57500.45 & $H$          & ...  & ...  & ...  & ...  & 3.8  & 4.8  & 4.8  & 4.7 & 5.8 & 6.4 & 8.1 & 9.1 & 9.8 & ...  & TW  \\
204864076 & 54249.5  & $K_{\rm S}$          & ...  & 2.86 & 4.45 & 4.94 & 4.94 & 4.86 & ...  & ... & ... & ... & ... & ... & ... & ... & K08$^{c}$ \\
205080616 & 54256.5  & $K_{\rm p}$+NRM  & 3.79 & 5.64 & 6.46 & 6.35 & 6.35 & 6.20 & 5.77 & 5.77 & ... & ... & ... & ... & ... & ... & K08$^{c}$ \\
205151387 & 54256.5  & $K_{\rm p}$ +NRM & 3.81 & 5.63 & 6.38 & 6.33 & 6.33 & 6.14 & 5.72 & 5.72 & ... & ... & ... & ... & ... & ... & K08$^{c}$ \\ 
205238942 & 57225.29 & $K_{\rm p}$+NRM+C600 & 0.55 & 3.59 & 4.49 & 4.23 & 4.9  & 5.9  & 6.6  & 7.1  & 7.4 & 8.0 & 9.2 & 10.7 & 11.4 & 11.8 & TW \\
205345560 & ...      & ...          & ...  & ...  & ...  & ...  & ...  & ...  & ...  & ... & ... & ... & ... & ... & ... & ... & ... \\
\hline
\end{tabular}
\begin{tablenotes}
\item[a] B19$=$\cite{Barenfeld2019}; C15$=$\cite{Cheetham2015}; G05$=$ \cite{Grady2005}; K08$=$\cite{Kraus2008}; M10$=$\cite{McClure2010}; R05$=$\cite{Ratzka2005}; TW = This Work. 
\item[b] \cite{Ratzka2005} and \cite{McClure2010} did not report detection limits for sources with detected candidate companions (see Table~\ref{tab:ao}). \cite{Grady2005} did not report detection limits.
\item[c] \cite{Cheetham2015}, \cite{Kraus2008} used nonredundant aperture masking (see Section~\ref{sec:ao}) and reported limits at separations of 10--20, 20--40, 40--80, 80--160, 160--240, 240--320~mas. \cite{Barenfeld2019} combine nonredundant aperture masking (using these same separation regions) with normal imaging (with limits reported at 40--80, 80--160, 160--240, 240--320, 320--500, 500--1000, and $>$1000~mas) and here we report the highest achieved contrast when the separation regions overlap.
\item[d] NRM$=$non-redundant aperture masking, C600$=$Coronograph
\end{tablenotes}
\end{threeparttable}
\end{table}
\end{landscape}
}

Table~\ref{tab:contrasts} presents the detection limits derived from high-contrast imaging of the dippers in our sample, when available. The derivation of the contrast limits for the sources observed as part of this work are described in Section~\ref{sec:ao} and we provide references for those values taken from the literature. Some sources have more than one entry when different works probed different separations.

\section{MCFOST Model Grid}
\label{appendix-B}

Our MCFOST models \citep{Pinte2006} all assume axisymmetric, tapered-edge disc profiles that are one of the standard options of the software package. We generate the model grid by varying six parameters: dust mass ($M_{\rm dust}=1\times10^{-6}$, $1\times10^{-5}$, $1\times10^{-4}$, $1\times10^{-3}$~$M_\odot$), critical radius ($R_c=10$, 30, 100~au), flaring exponent ($f_{\rm exp}=1.0$, 1.15, 1.30), scale height ($H_0=5$, 10, 15, 20~au at 100~au radius), surface density exponent ($\gamma=0.0$, 0.5, 1.0, 1.5), and maximum grain size ($a_{\rm max}=10$, 100, 1000, 10000~$\mu$m). To better represent empirical distributions of these model parameters, we weight $M_{\rm dust}$ according to the distribution in Taurus, and weight $H_0$ based on the distribution of midplane temperatures across the entire grid, which drastically reduces the $H_0=20$~au models as few discs are this hot. We also fix the stellar effective temperature ($T_{\rm eff}=4000$~K), inner radius ($R_{\rm in}=0.1$~au), and minimum grain size ($a_{\rm min}=0.01$~$\mu$m), while assuming Mie scattering and astronomical silicates \cite[similar to ][]{DL1984} for the dust grain properties. 

For each combination of model parameters, we generate a spectral energy distribution (SED) at 15 different inclinations from $i_{\rm d}=45$--90\degrees{} spaced uniformly in cos($i_{\rm d}$), then compute the fraction of models at a given inclination for which the flux from the system is at least 40 times weaker than that of the host star. This factor of 40 is chosen based on the median {\it Kepler} magnitude of our sample of $K_{\rm p} \approx 13$~mag and the {\it Kepler} faint limit of $K_{\rm p} \approx 17$~mag. We find that $\gtrsim50$~percent of the model grid would be undetectable by {\it Kepler} for discs with $i_{\rm d}\gtrsim80$\degrees{}, a function that rises steeply as $\gtrsim90$~percent would be undetectable for $i_{\rm d}\gtrsim85$\degrees{}.

\section{Rejected Dippers with Inclinations Reported in Literature}
\label{appendix-C}

Several dippers in Upper Sco have disc inclinations derived from ALMA data reported in the literature but are not included in our sample due to their large uncertainties. Indeed, the visibilities of all of these sources are flat or nearly flat with UV distance \cite[see Figure~1 in][]{Barenfeld2017}, indicating they are unresolved or poorly resolved. This make them distinct from the sources in our sample, which all have clearly declining visibilities with UV distance (see Figure~\ref{fig:data}). The excluded sources are given in Table~\ref{tab:rejects}, which has the same column header meanings and references as Table~\ref{tab:results}. For most of these sources we cannot derive any inclination using our method (Section~\ref{sec:discinc}), however for two discs we could extract inclinations consistent with those in the literature, albeit with large ($\gtrsim$20\degrees{}) uncertainties (using the archival data from 2013.1.00395.S for EPIC~204278916 and data from our targeted ALMA programme 2016.1.00336.S for EPIC~204757338) as given in Table~\ref{tab:rejects}. A handful of $\rho$~Oph dippers also have marginally resolved discs with large ($\gtrsim20$\degrees{}) inclination uncertainties when fit with our method; these are also not included in our sample.

We do not include these marginally resolved sources in our sample because the key algorithm implemented in {\tt emcee} has difficulties handling multi-modal posterior distributions, such as those of the inclination parameter due to its symmetry (i.e., $i_{\rm d}=\pm 20$\degrees{} are equally reasonable solutions), when the source is only marginally resolved and/or the signal-to-noise ratio is low. This is because the walker-based algorithm will still explore both posterior peaks, even if the priors are constrained to one period (i.e., $i_{\rm d} =$ [0\degrees{},90\degrees{}]), causing tails in the distribution that are artifacts of the algorithm rather than reliable posterior samples.

\begin{figure}
\begin{center}
\includegraphics[width=8cm]{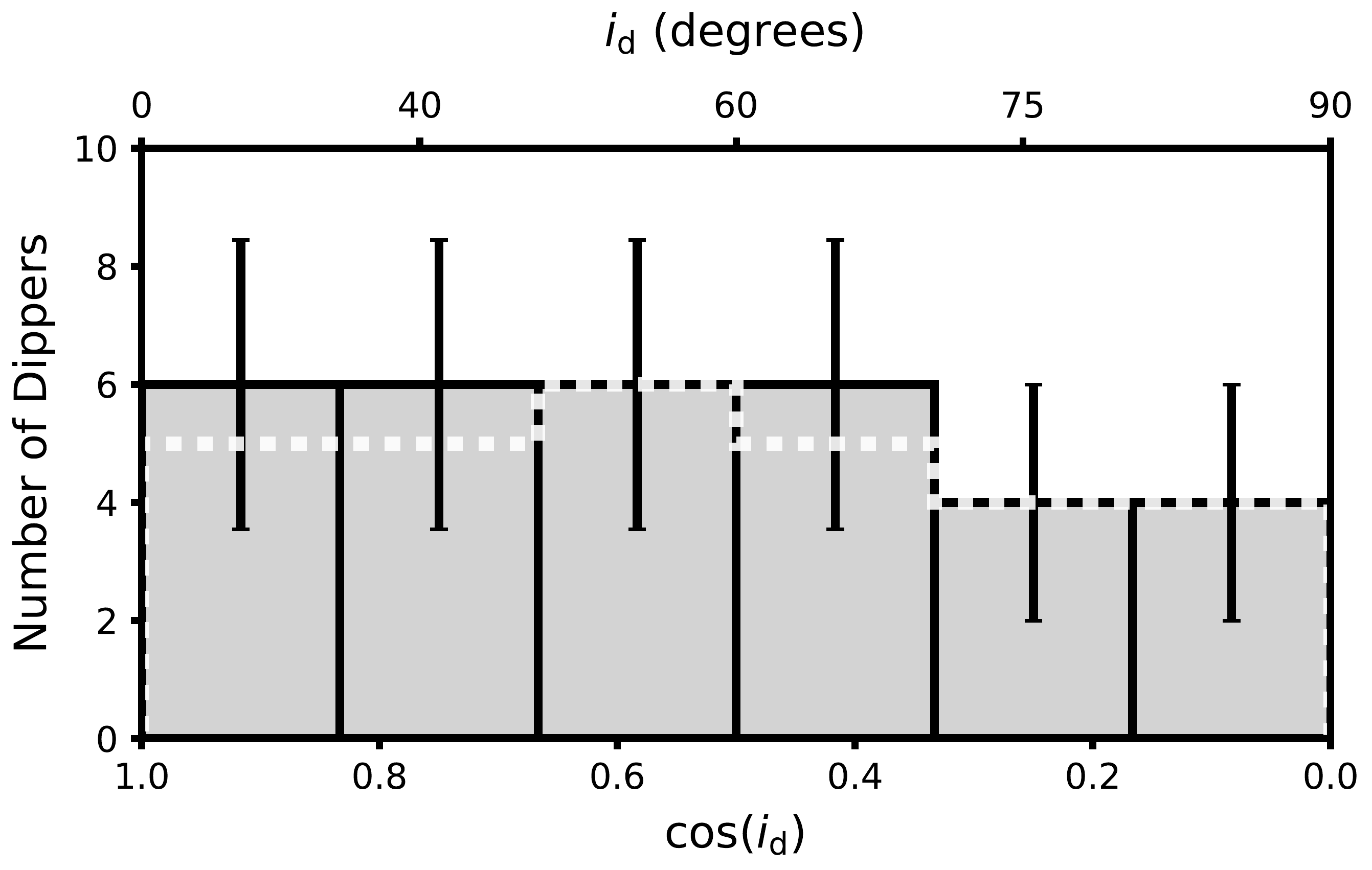}
\caption{The inclination distribution of dipper discs resolved by ALMA, now including those with large uncertainties from the literature that were used in Cody \& Hillenbrand (2018) to suggest a bias towards highly inclined discs amongst the dippers. The disc inclination values were taken from Table~\ref{tab:results} and Table~\ref{tab:rejects}, and the symbols are the same as in Figure~\ref{fig:inc_dist}.}
\label{fig:inc_dist_rejects}
\end{center}
\end{figure}

Nevertheless, including the sources in Table~\ref{tab:rejects} with the inclinations reported in the literature would not change our overall results. In fact, if taken at face value, they would actually make the distribution even more consistent with isotropic by filling out the higher inclinations, as shown in the histogram in Figure~\ref{fig:inc_dist_rejects}. We do not construct an ECDF as in Section~\ref{sec:incdist} because the MCMC posterior distributions are unreliable, for the reasons described above.

\begin{table}
\centering
\caption{Rejected Dippers with Inclinations Reported in Literature}
\label{tab:rejects}
\renewcommand*{\arraystretch}{1.4} 
\begin{tabular}{lccrrr} 
\hline
EPIC & SpT. & Ref. SpT & $i_{\rm d, lit}$ & Ref. $i_{\rm d, lit}$ & $i_{\rm d}$ \\
\hline
203750883  & M3.5 & L12 &  $86^{+4}_{-52}$  & B17 & ... \\
204245509  & K2   & L12 &   $4^{+48}_{-3}$  & B17 & ... \\ 
204278916  & M1   & L12 &  $57^{+14}_{-19}$ & B17 & $51^{+14}_{-17}$ \\
204757338  & M4.5 & A16 &  $68^{+10}_{-49}$ & B17 & $56^{+18}_{-27}$ \\
204932990  & M3.5 & L12 &  $86^{+4}_{-42}$  & B17 & ... \\
205037578  & M3.5 & L12 &  $80^{+7}_{-50}$  & B17 & ... \\ 
205241182  & M4.5 & L12 &  $71^{+8}_{-63}$  & B17 & ... \\  
205383125  & M3   & L12 &  $86^{+4}_{-60}$  & B17 & ... \\  
\hline
\end{tabular}
\end{table}

\bsp	
\label{lastpage}
\end{document}